\documentclass[conference]{IEEEtran}
\IEEEoverridecommandlockouts
\usepackage{amsmath,amsfonts}
\usepackage{algorithmic}
\usepackage{graphicx}
\usepackage{textcomp}
\usepackage{xcolor, soul}
\usepackage{url}
\usepackage{fancyhdr}
\usepackage{textcomp}
\usepackage{tabularx,ragged2e}
\usepackage{booktabs} %
\usepackage{fancyhdr}
\usepackage{amsfonts}       %
\usepackage{nicefrac}       %
\usepackage{microtype}      %
\usepackage[normalem]{ulem}
\usepackage{multirow}
\useunder{\uline}{\ul}{}
\usepackage[labelfont=bf, skip=0pt]{caption}
\usepackage{graphics}
\usepackage{xcolor, colortbl}
\usepackage{multicol}
\usepackage{physics}
\usepackage{graphicx}
\usepackage[labelformat=simple]{subcaption}

\usepackage{makecell}
\usepackage{float}
\usepackage{mathtools, nccmath}
\usepackage[plainruled,vlined,linesnumbered,noresetcount]{algorithm2e}
\usepackage{diagbox}
\usepackage{footmisc}
\usepackage{gensymb}
\usepackage{adjustbox}
\usepackage{mwe}
\usepackage{pifont}%
\usepackage{cite}
\usepackage{hyperref}
\usepackage{etoolbox}

\usepackage{tikz}

\makeatletter
\patchcmd{\@algocf@start}%
  {-1.5em}%
  {0pt}%
  {}{}%
\makeatother

\makeatletter
\newcommand{\removelatexerror}{\let\@latex@error\@gobble}
\makeatother

\newcommand{\countervalue}{counter value}
\newcommand{\countervalues}{counter values}

\newcommand{\versionnumber}{version number}
\newcommand{\versionnumbers}{version numbers}
\newcommand{\VN}{VN}
\newcommand{\VNs}{VNs}

\newcommand{\mpt}{MGX}

\usepackage{tabu}
\usepackage{cases}
\usepackage{comment}

\def\BibTeX{{\rm B\kern-.05em{\sc i\kern-.025em b}\kern-.08em
    T\kern-.1667em\lower.7ex\hbox{E}\kern-.125emX}}
\begin{document}

\title{MGX: Near-Zero Overhead Memory Protection for Data-Intensive Accelerators}

\author{
\IEEEauthorblockN{Weizhe Hua$^{\dagger}$, Muhammad Umar$^{\dagger}$, Zhiru Zhang$^{\dagger}$, G. Edward Suh$^{\dagger\mathsection*}$}
\thanks{$^*$Work was done while at Cornell University.}
\IEEEauthorblockA{\textit{$^{\dagger}$Cornell University, Ithaca, NY, USA}\\{\textit{$^{\mathsection}$Meta AI, Cambridge, MA, USA}}}\\
\textit{\{wh399,mu94,zhiruz,gs272\}@cornell.edu, edsuh@fb.com}
}

\maketitle

\begin{abstract}
This paper introduces MGX, a near-zero overhead memory protection scheme for hardware accelerators. 
MGX minimizes the performance overhead of off-chip memory encryption and integrity verification by exploiting the application-specific properties of the accelerator execution.
In particular, accelerators tend to explicitly manage data movement between on-chip and off-chip memories. 
Therefore, the general memory access pattern of an accelerator can largely be determined for a given application.
Exploiting these characteristics, MGX generates version numbers used in memory encryption 
and integrity verification using on-chip accelerator state rather than storing them in the off-chip memory; it 
also customizes the granularity of the memory protection to match the granularity used by 
the accelerator.
To demonstrate the efficacy of MGX, we present an in-depth study of MGX for DNN and graph algorithms.
Experimental results show that on average, MGX lowers the performance overhead of memory protection from 28\% and 33\% to 4\% and 5\% for DNN and graph processing accelerators in a wide range of benchmarks, respectively. 
\end{abstract}

\section{Introduction}
As the technology scaling slows down, computing systems are increasingly relying
on hardware accelerators to improve performance and energy efficiency.
For example, modern machine learning (ML) models such as deep neural networks (DNNs) are often quite compute-intensive and increasingly run on hardware accelerators~\cite{google2017tpu, microsoft2018brainwave} for both performance and energy efficiency.
Similarly, hardware accelerators are widely used for other compute-intensive workloads such as 
video decoding, signal processing, cryptographic operations, genome assembly, etc. 
This paper proposes a novel off-chip memory protection scheme for hardware accelerators, named
MGX (\textbf{M}emory \textbf{G}uard for \textbf{X}elerators), using secure acceleration of DNN and graph algorithms as the primary applications.

In many applications, the hardware accelerators may process private or sensitive data, which need strong security protection.
For example, ML algorithms often require collecting, storing, and processing a large amount of personal and potentially private data from users to train a model. 
Moreover, due to its high computational demand, both training and inference are often performed on a remote server rather than a client device such as a smartphone, implying that the private data and ML models may be exposed if the server is compromised or malicious.

A promising approach to providing strong confidentiality and integrity guarantees under untrusted environments is to create a hardware-protected trusted execution environment (TEE), also called an enclave as in Intel SGX \cite{sgx}. 
The cryptographic protection of off-chip memory represents an essential technology to enable the hardware-protected TEE.
The off-chip memory protection also represents the main source of performance overhead in the traditional secure processor designs.
For a general-purpose processor, the memory protection schemes need to be able to
handle any sequence of memory accesses to arbitrary memory locations, and typically protect memory accesses at a cache-block granularity. 
In secure processors, the counter-mode encryption is used to hide decryption latency, where the \countervalue~is typically a concatenation of the memory address and a \versionnumber~(\VN).
The \versionnumber~is stored in memory and incremented on each write of an encrypted block. 
To protect integrity of off-chip memory, a message authentication code (MAC) needs to be attached to each cache block in memory.
Moreover, the integrity verification requires a tree of MACs to prevent replay attacks.
Unfortunately, the \VN~and MAC accesses can lead to non-trivial performance overhead for memory-intensive workloads.
In order to extend the TEE to application-specific accelerators, we need a more efficient memory protection scheme that can protect memory-intensive workloads with low overhead.

In this paper, we show that memory encryption and integrity verification can be performed with almost
no performance overhead for an accelerator by customizing protection to 
the accelerator-specific memory access pattern.
We make key observations that the application-specific accelerators 
explicitly move data between on- and off-chip memories following a relatively
simple pattern that is specific to an application, and that the off-chip data movements usually use a granularity that is larger than a cache block. 
The relatively simple and static memory access patterns imply that
\versionnumbers~can often be calculated from the on-chip state without
storing them in off-chip memory. 
The coarse-granularity data movement suggests that the \versionnumbers~for memory encryption and
the MACs for integrity verification can be maintained at a coarse granularity to reduce the overhead.

We study the memory access behaviors of DNN inference and training as well as two representative graph algorithms, and show how to determine the \versionnumbers~using the scheduling and the state of the accelerator.
By generating \versionnumbers~on-chip and performing protection at an application-specific granularity,
MGX can eliminate most of overhead for off-chip memory protection; no \versionnumber~is stored in the off-chip memory, no integrity tree is needed, and each MAC protects a large amount of data instead of one cache block.

To evaluate the effectiveness of MGX for DNN accelerators, we performed extensive experiments using cycle-level simulations based on SCALE-Sim~\cite{scale-sim1}, 
an open-source DNN accelerator simulator from ARM research.
The overhead of applying MGX to both DNN inference and training is less than 5\% on the state-of-the-art DNN models.
For graph accelerator, we performed the experiments using a combination of RTL and cycle-level simulations based on an open-source graph accelerator~\cite{graphlily}.
The simulation results on two important graph algorithms, PageRank and Breadth-first Search (BFS), show that MGX can provide both memory encryption and integrity verification with very low overhead. 

This paper makes the following major contributions:
\vspace{-0.2em}
 \begin{itemize}
  \item[$\bullet$] We propose MGX, a near-zero overhead memory protection scheme for accelerators. MGX minimizes the performance overhead of memory protection by assigning \countervalues~for data using on-chip state and performing coarse-grained memory protection.   
  
  \item[$\bullet$] We demonstrate the applicability of MGX by showing a concrete implementation of MGX for DNNs and graph algorithms
  and detailed analyses of a genome sequence alignment accelerator and an H.264 video decoder accelerator. 
  
  \item[$\bullet$] We evaluate the secure accelerators with MGX and show that the overhead is 3.2\% and 4.7\% for DNN inference and training, and 5.1\% and 4.9\% for PageRank and BFS.
  
\end{itemize}
\section{Secure Accelerator Architecture}
\label{sec:arch}
\begin{figure}[!t]
  \begin{center}
        \includegraphics[scale=0.42]{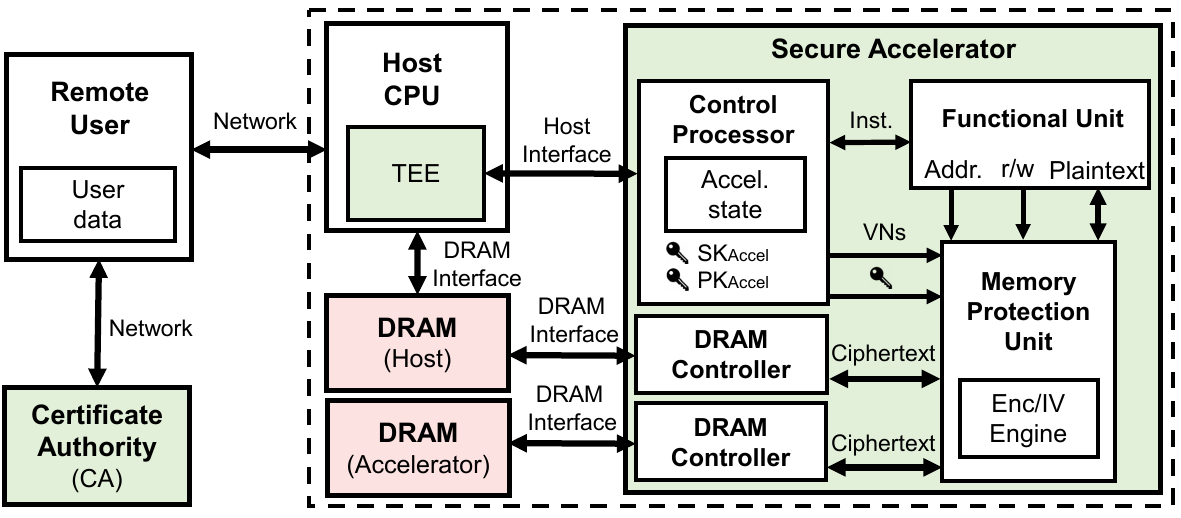}
        \vspace{0.1cm}
        \caption{Secure accelerator architecture overview --- \small{The green and red boxes represent trusted and untrusted components, respectively.}}
        \vspace{-0.3cm}
   \label{fig:arch}
  \end{center}
\end{figure}

The goal of a secure accelerator is to protect the confidentiality and the integrity of the state and data of the accelerator even in an untrusted environment where the host software and the off-chip memory cannot be trusted.
For example, the secure DNN accelerator aims to protect inputs, outputs, weights, and all intermediate results.

\begin{table}[!t]
\caption{Threats and defense in the secure accelerator.}
\resizebox{\columnwidth}{!}{
\begin{tabular}{@{}lll@{}}
\toprule
Threats & Defense & Mechanism \\ \midrule
\begin{tabular}[c]{@{}l@{}}Unauthorized access by \\ privileged process on host CPU\end{tabular} & CPU Enclave & Isolation using a CPU TEE \\ \midrule
\begin{tabular}[c]{@{}l@{}}Shared access to \\ the off-chip memory\end{tabular} & \begin{tabular}[c]{@{}l@{}}Off-chip Mem.\\ Protection\end{tabular} & \begin{tabular}[c]{@{}l@{}}Memory encryption and integrity\\ verification with MGX\end{tabular} \\ \midrule
Side-channel attacks & / & Not considered \\ \bottomrule
\begin{tabular}[c]{@{}l@{}}Corrupt the kernels running in\\ the accelerator\end{tabular} & Attestation & \begin{tabular}[c]{@{}l@{}}Kernels are attested and running \\ on the on-chip control processor\end{tabular} \\ \midrule
\begin{tabular}[c]{@{}l@{}}Communication channel between \\ the host and the accelerator\end{tabular} & Key Exchange & DHE key-exchange protocol \\ \midrule
\end{tabular}}
\label{tab:TCB}
\end{table}

Figure~\ref{fig:arch} shows the threat model of the secure accelerator and 
Table~\ref{tab:TCB} summarizes the potential threats and the corresponding protection mechanisms.
As shown in the figure, the hardware TCB mainly includes the CPU TEE and the accelerator.
The host processor is assumed to have a trustworthy TEE with a secure communication channel to the accelerator as in recent proposals for GPU TEEs \cite{graviton}.
Following the typical threat model for secure processors, we assume that the internal operations and state of an accelerator cannot be directly observed or modified by an adversary through physical attacks.
The off-chip memory is assumed untrusted;
the secure accelerator needs memory protection to encrypt confidential data and detect unauthorized changes in values stored in DRAM.
We do not consider side-channel attacks such as the memory address, timing, and power side channels.

The accelerator-specific kernels are attested and then executed on the trusted control processor of the accelerator.
For external communications, the accelerator needs to support a secure key-exchange protocol to establish trust and securely communicate with a remote user or a TEE.
Specifically, the secure accelerator includes a unique private key (SK$_\text{Accel}$), embedded by a manufacturer. 
We assume that a user obtains the corresponding public key using a private key infrastructure as in
Intel SGX or Trusted Platform Modules (TPMs).
The accelerator also provides remote attestation using its private key so that a user can authenticate hardware, the hash of firmware/configuration of the accelerator, the hash of the application kernel, and the hash of the input and output data.

Our threat model is representative of the typical TEE threat model
and common for both the baseline memory protection and MGX.
In MGX, both a memory protection unit and an application kernel that issues commands to the accelerator need to be trusted for memory protection.
The kernel also needs to be included in remote attestation and protected by running on a control processor inside
an accelerator using on-chip memory.
Note that software inside a TEE such as the application kernel is already
inside the TCB of a typical TEE; software inside the TEE is allowed to 
output a secret.

To use the secure accelerator, a user sends a command to initiate a new session, which will have the accelerator clear its internal state, set a pair of new symmetric keys for encryption and integrity verification, enable protection mechanisms, and establish a secure (encrypted and authenticated) communication channel with the user using a standard protocol such as an TLS.
After initialization, a user sends an application kernel and user data through the encrypted channel.
The accelerator loads the data by decrypting it with the communication key, and placing it in
protected memory that is encrypted with the memory encryption key.
Once the execution is finished, the accelerator returns the encrypted results.
\section{MGX: Near-zero overhead memory protection for accelerator}
\label{sec:low_overhead_mp}
This section first describes state-of-the-art memory protection scheme (i.e., memory encryption and integrity verification).
Then, we introduce the MGX scheme, which provides secure and low-overhead memory protection by leveraging the regular and mostly static memory access patterns of specialized accelerators.
Finally, we provide an example on tiled matrix multiplication to better explain the proposed MGX scheme.
\subsection{Memory Protection Basics}
\begin{figure}[t]
\begin{center}
\subfloat[Traditional protection.]{
\begin{minipage}{.7\linewidth}
    \centering
        \includegraphics[width=\columnwidth]{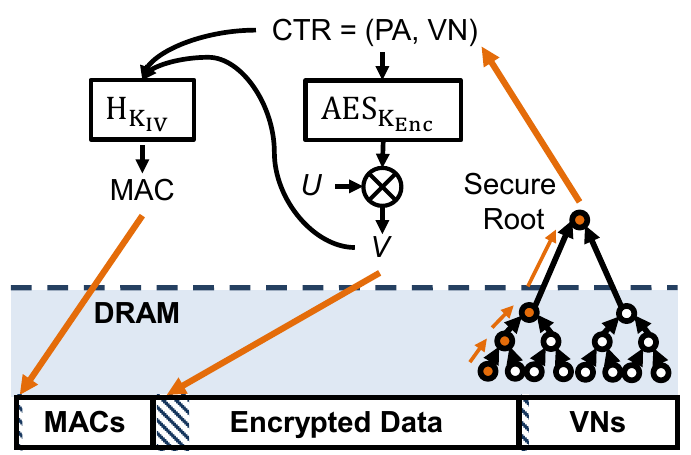}
        \vspace{-0.5cm}
        \label{fig:enc_iv}
\end{minipage}}
\hspace{0.02cm}
\subfloat[Proposed protection.]{
\begin{minipage}{.55\linewidth}
    \centering 
        \includegraphics[width=\columnwidth]{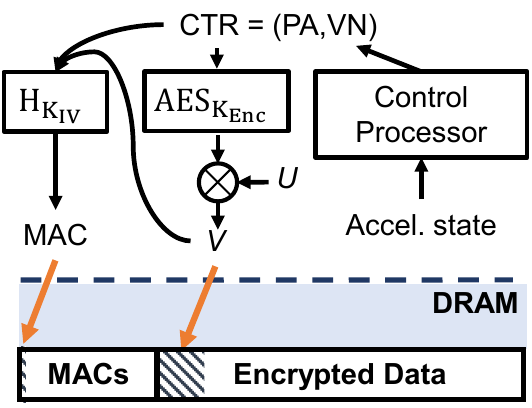}
        \vspace{-0.5cm}
        \label{fig:mgx_enc_iv}
\end{minipage}}
\end{center}
\caption{Memory encryption and integrity verification.}
\vspace{-0.25cm}
\end{figure}

\textbf{Memory Encryption.}
As shown in Figure~\ref{fig:enc_iv}, existing techniques~\cite{encrypt_survey, cpu_MEE, aegis} typically use the counter-mode encryption (AES-CTR) to hide AES latency.
AES-CTR requires a non-repeating counter value for each encryption under the same AES key.
In a secure processor, the \countervalue~often consists of the physical memory address (PA) of the data block that will be encrypted
and a (per-block) \versionnumber\ (VN) that is incremented on each memory write.
When a data block is written, the encryption engine increments the VN and then encrypts the data. 
When a data block is read, the encryption engine retrieves the VN used for encryption and then decrypts the block. 
Let K$_\text{Enc}$, $U$, $V$ be the AES encryption key, plaintext, and ciphertext, respectively. 
The AES encryption can be formulated as $V = U \oplus \text{AES}_{\text{K}_\text{Enc}}(\text{PA} || \text{VN})$, where $||$ and $\oplus$ represent bit-field concatenation and XOR, respectively.

As general-purpose processors can have an arbitrary memory access pattern, the VN for each cache block can be any value at a given time. 
In order to determine the VN for a later read, a secure processor needs to store the VNs in DRAM.
To avoid re-using the same \countervalue, the AES key needs to change once the VN reaches its maximum, which implies that the size of the VN needs to be
large enough to avoid frequent re-encryption.
For example, Intel SGX~\cite{cpu_MEE} uses a 56-bit VN for each 64-byte data block, which introduces 11\% storage and bandwidth overhead.
In general, the VNs cannot fit on-chip and are stored in DRAM.
As the VNs are stored in the off-chip memory, the integrity and freshness of VNs also need to be protected with MACs to ensure the confidentiality.

\textbf{Integrity Verification.}
To prevent off-chip data from being altered by an attacker, integrity verification cryptographically 
checks if the value from DRAM is the most recent value written to the address by the processor. 
For this purpose, a MAC of the data value, the memory address, and the VN is computed and stored
for each data block on a write, and checked on a read from DRAM.
However, only checking the MAC cannot guarantee the freshness of the data; a replay attack can
replace the data and the corresponding \VN~and MAC in memory with stale values without being
detected. 
To defeat the replay attack, a Merkle tree (i.e., hash tree) \cite{hash_tree} is used to
verify the MACs hierarchically in a way that the root of the tree is stored on-chip. 
As shown in Figure~\ref{fig:enc_iv}, a state-of-the-art method~\cite{iv_micro07} uses a Merkle tree to protect
the integrity of the VNs~in memory, and includes a \VN~in a MAC to ensure the freshness of data. 
Let us denote the key, plaintext, and ciphertext as K$_\text{IV}, U, V$, respectively. 
The MAC of an encrypted data block can be calculated as $\text{MAC} = \text{H}_{\text{K}_\text{IV}}(V||\text{PA}||\text{VN})$.
The overhead of integrity verification is nontrivial as it requires traversing the tree stored in DRAM.
To mitigate this overhead, recently verified MACs are stored in a cache.
However, caching is often not effective for data-intensive applications such as large ML models.
Moreover, the Merkle tree poses a scalability challenge because its depth needs to increase with the size of the  protected memory. 

\begin{figure}[t!]
\centering
\includegraphics[scale=0.71]{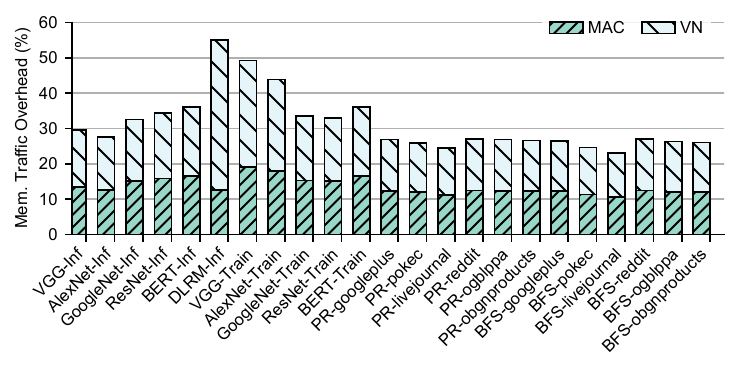}
\caption{Breakdown of the memory traffic overhead introduced by the traditional off-chip memory protection scheme --- \small{MAC and VN represent the overhead incurred by accessing MACs and VNs, respectively. Inf and Train stand for DNN inference and training tasks, respectively. PR and BFS stand for PageRank and Breadth-first Search, respectively}.}
\label{fig:bw_breakdown}
\end{figure}

Figure~\ref{fig:bw_breakdown} shows
the memory traffic overhead of applying the traditional memory protection for DNN inference and training, PageRank, and BFS.
For all applications, the number of memory accesses increases by at least 23.1\% and over 49.2\% in the worst case.
The average memory traffic overheads for DNN inference, DNN training, PageRank, and BFS are 36.1\%, 40.4\%, 26.3\%, and 25.6\%.
Among the applications, DNN training is most heavily affected by off-chip memory protection because it requires access to a large amount of data.
In addition, the memory traffic overhead of accessing the VNs can greatly exceed that of accessing the MAC, as it requires a Merkle tree to verify the integrity of the VNs.

\subsection{Intuition}
\label{sec:intuition}

The main overhead of memory protection comes from storing and accessing VNs and MACs for protecting the confidentiality and integrity of data and verifying the MACs for VNs hierarchically in the off-chip memory.
Because many accelerators such as DNN and graph accelerators are often memory-intensive, these additional meta-data accesses can lead to nontrivial performance overhead.
We propose to significantly reduce the memory protection overhead by generating \VNs~without storing them in memory and
customizing the protection granularity based on the application-specific memory access pattern.

As customized for a particular application domain, 
specialized accelerators tend to have more predictable and regular memory access patterns compared to general-purpose CPUs.
In particular, both DNN inference and training can be scheduled statically based on the network structure.
For example, most popular deep learning (DL) frameworks such as TensorFlow~\cite{tensorflow} adopt declarative programming and lazy execution, where the DNN network is represented as a data-flow graph (DFG).
A DL framework (i.e., compiler) first optimizes the DFG and then generates the scheduling of the computations and the corresponding memory accesses based on the optimized DFG before execution.
Therefore, the DL compiler can assign a VN for each memory access based on the schedule without storing the VNs in memory.

Moreover, accelerators may have the same access pattern to a large chunk of memory because they typically operate on blocks of memory larger than cache lines.
For example, DNNs write the output feature maps of a layer to DRAM the same number of times because each output feature map is generated following the same schedule.
As VNs reflect the maximum number of writes to the corresponding memory block, this regular memory access pattern means that we only need one VN for all the output features of a layer or a tile.
If a DNN accelerator only writes the output features to DRAM once per layer (i.e., no tiling), MGX can simply use the layer number as part of the VN.
In addition to DNN accelerators, most graph accelerators update the attributes associated with each vertex the same number of times (e.g., mostly once) in each iteration.
In that case, the attribute values of vertices can also share the same VN value.

\newcommand{\matmul}{%
\begingroup
\removelatexerror%
\begin{algorithm}[H]
\footnotesize{
\SetAlgoLined
\SetKwInOut{Input}{Input}
\SetKwInOut{Output}{Output}
\Input{Tiles of Matrix A ($A_{1-2}$) \\ Tiles of Matrix B ($B_{1-4}$)}
\Output{Matrix C (tiles $C_{1-2}$)}
Init.: VN$[A] = $ VN$[B] = $ VN$[C] = n$; $C = 0$ \\
{
\For{$i = 1;\ i <= 2;\ i\mathrel{+}=1$}{
    \For{$j = 1; \ j <= 2; \ j\mathrel{+}=1$}{
        $C_j \mathrel{+}= A_i\cdot B_{j+2*(i-1)}$
    }
    \tcp{Write $C$ with the incremented VN.}
    VN[$C$] += 1
}
}
}
\end{algorithm}
\endgroup
}

\begin{figure*}
\begin{minipage}{\linewidth}
\centering
 \begin{subfigure}{.25\textwidth}
 \includegraphics[width=\linewidth]{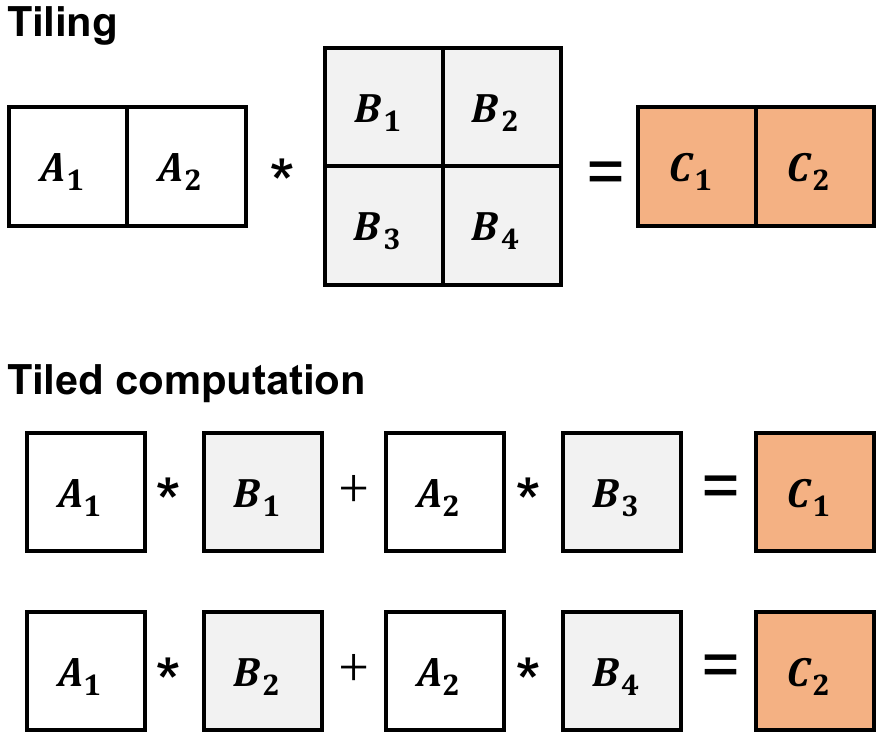}
  \captionof{figure}{Illustration of a tiled \texttt{MatMul} between two matrices $A$ and $B$ --- \small{the input A and B are partitioned into two ($A_1$ and $A_2$) and four ($B_1, ..., B_3$) tiles} respectively.}
  \label{fig:matmul_example}
 \end{subfigure}
 \hfill
 \begin{subfigure}{.3\textwidth}
 \matmul
 \captionof{figure}{Pseudocode of the tiled \texttt{MatMul} kernel --- \small{the inner loop ($i$) calculates the partial results of $C_1$ and $C_2$ and the outer loop ($j$) accumulates the partial results to get the final results}.}
 \label{list:matmul_schedule}
 \end{subfigure}
 \hfill
 \begin{subfigure}{.38\textwidth}
  \centering
  \includegraphics[width=0.95\linewidth, clip]{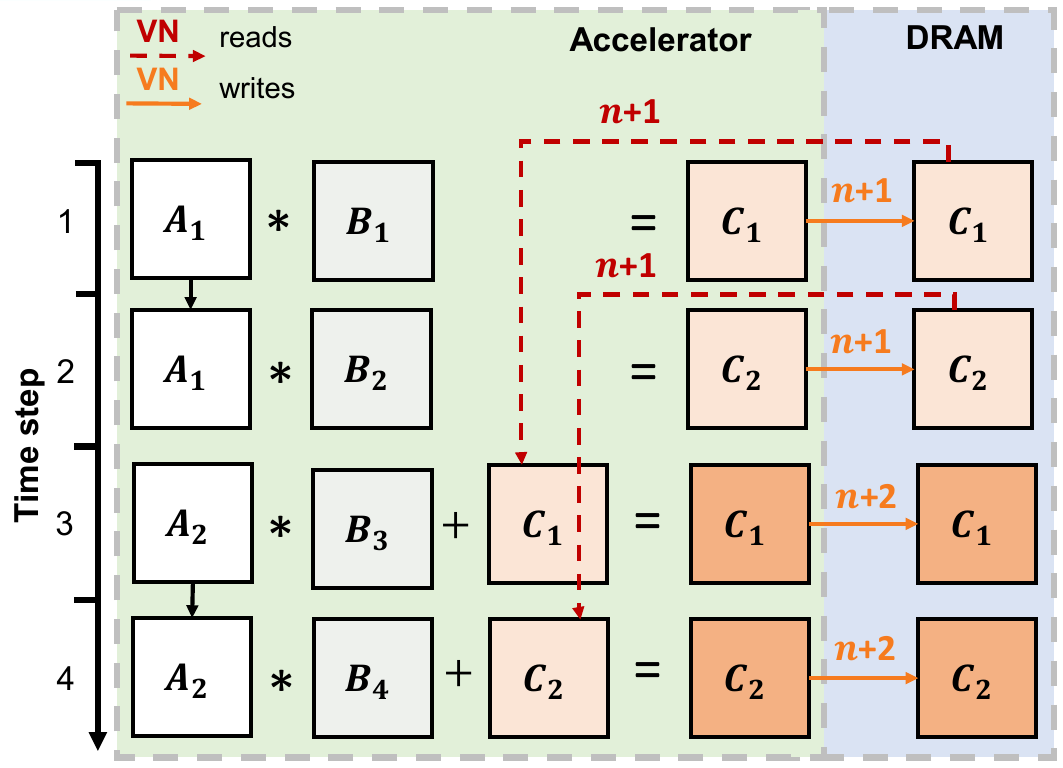}
  \captionof{figure}{VN generation scheme for the tiled \texttt{MatMul} kernel --- \small{Red values indicate the VNs for reads whereas orange values represent the VNs for writes. $A$ and $B$ are previously written to off-chip memory with VN value $n$. Memory reads to A and B are omitted.}}
  \label{fig:matmul_VN}
 \end{subfigure}
 \caption{Illustration of the VN generation scheme for the tiled \texttt{MatMul} example given a specific scheduling.}
 \label{fig:matmul}
\end{minipage}
\end{figure*}

\subsection{MGX Scheme}
\label{sec:mgx_scheme}
A specialized accelerator typically has an on-chip control processor that receives the statically compiled kernel (i.e., the accelerator-specific executable/binary) and
is responsible for executing the kernel, orchestrating the functional units of the accelerator, and keeping track of the state of the accelerator.
For example, the BFS algorithm can be considered as a kernel for graph accelerators.
In our MGX design, as shown in Figure~\ref{fig:mgx_enc_iv}, the VNs for reading and writing memory blocks are generated by the kernel running on the trusted on-chip control processor, rather than being stored in memory.
The application kernel can assign VNs for memory writes based on the scheduling of compute/memory operations and accelerator state.
We found that the kernel only needs to maintain an on-chip state to generate VNs, given an application-specific and coarse-grained nature of accelerator memory accesses.
For memory writes, the kernel ensures that the VN is greater than the last-used VN value for the memory location so that the same VN value is never reused for encrypting a memory block.
For memory reads, the kernel on the control processor regenerates the VN value used for the most recent write to the same address by using the on-chip state to ensure proper decryption.

In the DNN accelerator, a kernel on the control processor implements a full DNN model,
and VNs are computed inside the accelerator without any off-chip communications.
However, depending on the complexity of the application, an accelerator may rely more on the host CPU
to determine which operations to run.
In such cases, the scheduling software in the host CPU TEE can provide additional state to determine VNs 
when issuing commands to an accelerator control processor.
Note that the VN values can be public because the security of the AES-CTR encryption and the MAC only
depends on the integrity, not the confidentiality, of VNs.

It is worth noting that MGX does not require static or sequential memory access patterns to generate VNs. 
Reads do not affect the VNs no matter how irregular they are.
Writes can also happen in an arbitrary order using one VN value as long as they occur once per each address.
If needed, the control processor can keep additional state for VNs.

Once the VN is determined, the encryption (Enc) engine can decrypt/encrypt 
each 128-bit data block using the standard AES-CTR method for memory encryption.
As VNs no longer need to be stored in DRAM, the integrity protection tree for VNs also becomes unnecessary, greatly reducing off-chip accesses.
For integrity protection, MACs still need to be stored in memory. 
We propose to further reduce the overhead by customizing the size of a memory block that each MAC protects to match the data movement granularity of the accelerator.
For example, the CHaiDNN accelerator~\cite{chaidnn} from Xilinx reads a 512-byte chunk from memory at a time.
Using a 64-bit MAC for each 512-byte data block significantly reduces the bandwidth overhead for integrity protection.

Figure~\ref{fig:matmul_example} shows an example of tiled matrix multiplication (\texttt{MatMul}), where two matrices $A$ and $B$ are blocked into two and four tiles, respectively.
According to the scheduling shown in Algorithm~\ref{list:matmul_schedule}, the partial results of $C_1$ and $C_2$ are first computed; then the final results are obtained by summing the partial results.
We assume that $A$ and $B$ are previously written to off-chip memory with VN value $n$.
Since $A$ and $B$ are read-only during the computation, the {\tt MatMul} kernel uses $n$ as the VN for reading tiles of $A$ and $B$, as shown in Figure~\ref{fig:matmul_VN}.
In the first two iterations (i.e., iteration 1 and 2 in Figure~\ref{fig:matmul_example}, the accelerator writes the partial results of $C_1$ and $C_2$ to DRAM with an incremented VN value $n+1$.
The VN value for $C_1$ and $C_2$ needs to be incremented as the memory locations can be reused.
Since $C_1$ and $C_2$ occupy different memory locations in the off-chip memory under this specific scheduling, the {\tt MatMul} kernel can assign the same VN value $n+1$ for both tiles.
For the last two rounds, the {\tt MatMul} kernel first reads the partial results of $C_1$ and $C_2$ with VN value $n+1$, which is the VN value used to write them. 
Then, the kernel increments the VN and write the final results of $C_1$ and $C_2$ with VN value $n+2$.
The kernel keeps track of the VN for each matrix (or each tile) in its program state (i.e., VN[A], VN[B], VN[$C_1$], and VN[$C_2$].)

\subsection{Security Analysis}
\textbf{Encryption --}
MGX uses the same AES counter-mode encryption that is used by the traditional memory encryption scheme for processors.
The only difference between MGX and the traditional scheme is that the VN in MGX is determined/generated by the scheduler based on the accelerator state, rather than being stored in off-chip memory.
As long as the version number generator guarantees that the VN is unique for each write to a given memory location, the counter value, which is a concatenation of the memory address and the version number, is different for each encryption. %
Therefore, the security of the memory encryption in MGX can be reduced to the AES counter-mode encryption, which
is one of the approved modes of operation~\cite{aes_nist, aes_comment}.
Note that using one VN for multiple memory locations does not sacrifice security
because the counter value to a block cipher in the counter mode includes a memory address in addition to a VN. 

\smallskip
\textbf{Integrity Verification --}
MGX uses a MAC to protect the integrity of data in memory. The MAC includes the address and the VN
in addition to data.
This MAC construction is identical to the one that is used in the traditional integrity verification scheme
(shown in Figure~\ref{fig:enc_iv}), and protects against replay, relocation, and substitution attacks~\cite{hash_tree_proof},
as long as the version numbers are unique for each write to a location.
In the traditional scheme, the version numbers need to be protected separately using a Merkle tree because they are stored
in off-chip memory. 
In MGX, version numbers cannot be tampered by an attacker 
because they are generated on-chip.
Also, MGX requires the scheduler to generate VNs in a way that
a VN value is only used at most once for a write to each memory location.
Thus, the integrity protection in MGX can be reduced to that of the chosen keyed-hash function.
\section{MGX for DNN Acceleration}

\begin{figure*}
\centering
\begin{minipage}{\linewidth}
  \centering
  \includegraphics[width=0.85\linewidth]{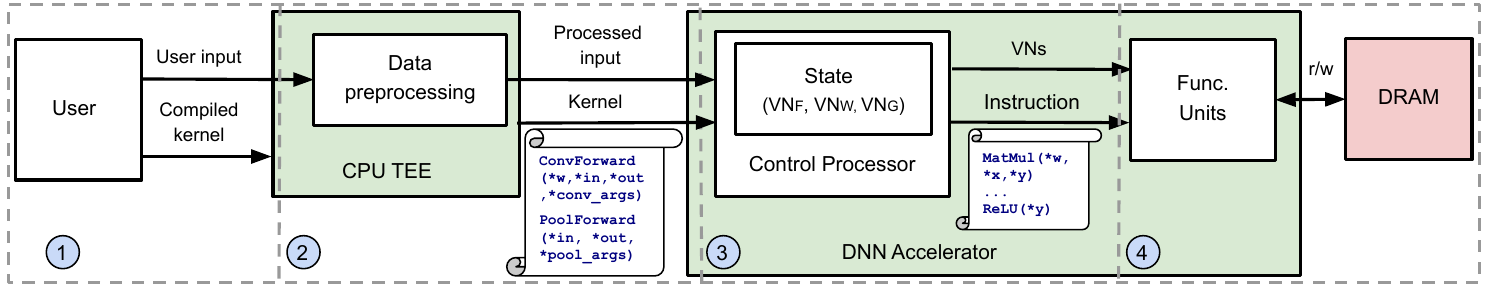}
  \captionof{figure}{The workflow of secure acceleration of DNNs with the proposed MGX scheme.}
  \label{fig:mgx_flow}
\end{minipage}%
\end{figure*}

This section introduces the background on DNNs, 
the workflow of DNN acceleration, and discusses how MGX can be applied %
to enable efficient memory encryption and integrity verification for secure DNN computation.

\subsection{Background on DNNs}
DNNs mainly consist of convolutional (conv), dense, normalization, activation, and pooling layers. 
The DNN inference is usually executed in a layer-by-layer fashion, where each layer takes either an external input (e.g., the first layer) or input features generated by the previous layer(s) to produce output features for the subsequent layer(s).
For each conv/dense layer, the DNN accelerator fetches the input features ($\mathbf{x}$) and weights ($\mathbf{w}$) from DRAM, generates the output features ($\mathbf{y}$) by computing $\mathbf{y} = \mathbf{w} * \mathbf{x}$, and stores the output features back to DRAM.
The DNN inference finishes after executing the last layer and generates a prediction. 

One iteration of DNN training consists of a forward propagation and a backpropagation. 
The forward pass is the same as the inference except that training requires computing the loss and the intermediate features need to be saved. 
After the loss is calculated, it is propagated in a backward manner through the entire network. 
For each layer, the DNN accelerator fetches the gradients from the subsequent layer ($\mathbf{g_y}$), reads input features ($\mathbf{x}$) and weights ($\mathbf{w}$) from off-chip memory, computes the gradients with respect to (w.r.t) the input features ($\mathbf{g_x} = \mathbf{g_y} * \mathbf{x}$) and weights ($\mathbf{g_w} = \mathbf{g_y} * \mathbf{w}$), updates the weights by calculating $\mathbf{w} \mathrel{+}= -\alpha\cdot\mathbf{g_w}$, where $\alpha$ is the learning rate, and stores $\mathbf{g_x}$ to the DRAM.
The gradients w.r.t the inputs ($\mathbf{g_x}$) are used as the output gradients ($\mathbf{g_y}$) for the previous layer.
The backpropagation continues until reaching the first layer.

\subsection{Workflow of Secure DNN Acceleration}
As depicted in Figure~\ref{fig:mgx_flow}, we can break down the workflow of DNN acceleration into four steps.
\textcircled{1} A user sends the private input and compiled kernel (i.e., DNN executable for the inference/training task) to the CPU enclave. 
We assume that the inference/training job is statically compiled into an accelerator-specific kernel by the user in the offline phase using a DNN compiler such as PyTorch Glow~\cite{glow} and TVM~\cite{tvm}.
Alternatively, the DNN compiler can be executed within the CPU TEE.
Since memory-related optimizations, such as instruction scheduling and static memory allocation, are performed at compilation time, all memory accesses are determined prior to execution.
\textcircled{2} The CPU TEE processes inputs (e.g., data augmentation for image data) and then forward the processed data and the kernel to the DNN accelerator.

{\textcircled{3} 
The on-chip control processor executes the kernel.
For accelerators that support high-level functions (e.g., convolution and pooling) such as TPU-v1~\cite{google2017tpu}, TVM-VTA~\cite{vta}, and CHaiDNN~\cite{chaidnn}, 
the high-level functions in the kernel can issue multiple low-level instructions to functional units.
For example, the \texttt{MatMul} instruction is executed by the matrix multiplication array on the accelerator.
The control processor is also responsible for providing the VN values for memory reads and writes required by each instruction.
For example, the convolution function
in some accelerators is implemented as a nested for loop, where the inner loop computes the partial results of different tiles of the output features and the outer loop accumulates the partial results to obtain the final results of the output features.
Similar to the \texttt{MatMul} example discussed in Section~\ref{sec:mgx_scheme}, the kernel code provides the VN values for each tile in the inner loop of the function.
{\textcircled{4}} Finally, the functional unit performs the DNN computation specified by the low-level instructions.}

\subsection{Version Number Generation for DNNs}
\label{sec:dnn_vn}

\begin{figure}
\centering
\begin{minipage}{\linewidth}
  \centering
  \includegraphics[width=\linewidth]{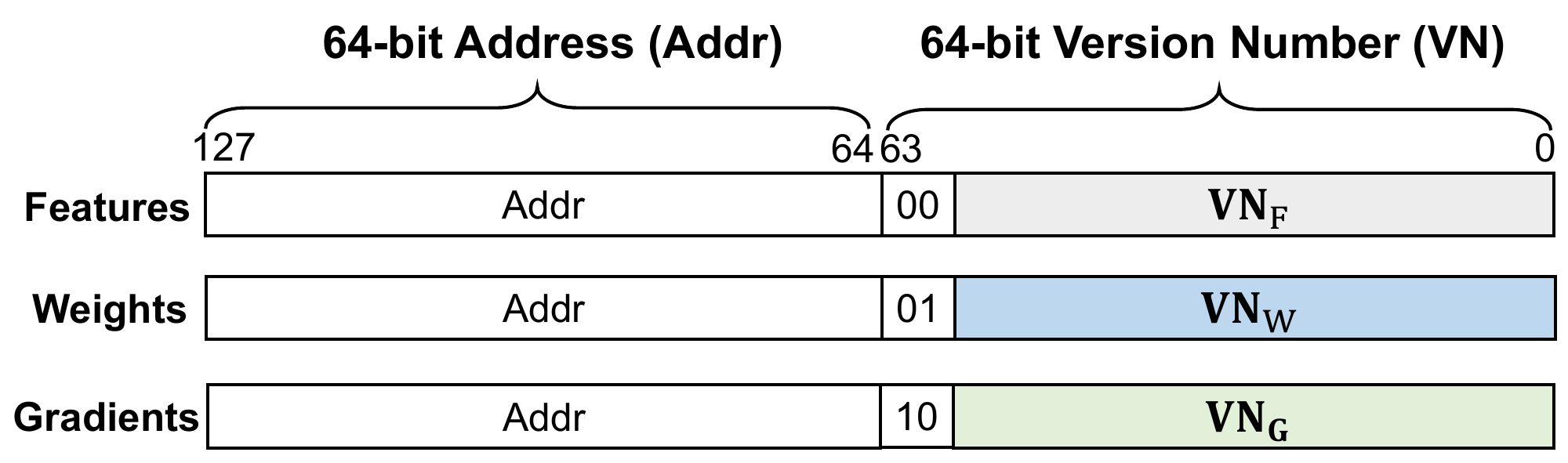}
  \vspace{-0.3cm}
  \captionof{figure}{Counter construction for DNN features, weights, and gradients.}
  \label{fig:counters}
\end{minipage}%
\vspace{-0.2cm}
\end{figure}

This subsection describes the \mpt~scheme for DNN inference and training, focusing on how VNs can be determined by the DNN kernel on the control processor.
Figure~\ref{fig:counters} shows how the counter values are constructed for DNN memory protection in MGX.
Each counter value includes the address of the memory block being encrypted/decrypted and a 64-bit VN,
and is used as the input to AES-CTR encryption.
Note that using one VN for multiple memory locations does not sacrifice security because the counter value includes a memory address.
The VNs are constructed differently for accessing three different types of data: features, weights, and gradients, ensuring that the counter values are unique for each encryption and are never reused.

\noindent\textbf{DNN Inference --}
During DNN inference, the accelerator reads the feature maps of the previous layer as the input, performs the computation, and then produces the output feature maps of the current layer.
The feature maps of each layer are written to off-chip memory the same number of times, regardless of the scheduling of the accelerator.
Therefore, we can keep one unique VN value for the feature maps of each layer (VN$_\text{F}$).
When writing new feature maps generated from a DNN layer, the DNN kernel increments the maximum VN$_\text{F}$ used before and assign this new value as the VN$_\text{F}$ for these feature maps.
In the case that the feature maps are written exactly once at the end of each layer, 
the DNN kernel can assign VN$_\text{F}$ for the feature maps based on the layer number.
For example, the feature maps of the $i^{th}$ layer have a VN$_\text{F}$ value of $i$ concatenated with the input count.

\newcommand{\dnntiling}{%
\begingroup
\removelatexerror%
\begin{algorithm}[H]
\footnotesize{
\SetAlgoLined
\SetKwInOut{Input}{Input}
\SetKwInOut{Output}{Output}
\Input{Input $x$ \& weight $w$}
\Output{Output $y$}
Init.: VN$_\text{F}[y]$ = VN$_\text{F}[x] = n$\\
{
\For{$i = 1;\ i \leq t;\ i\mathrel{+}=1$}{
\texttt{Read}($x_t$, VN$_\text{F}[x]$)\\
\If {$i > 1$}{
\texttt{Read}($y$, VN$_\text{F}[y]$)\\
}
$y \mathrel{+}= w_t * x_t$\\
VN$_\text{F}[y] \mathrel{+}=1$\\
\texttt{Write}($y$, VN$_\text{F}[y]$)\\
}
}
}
\end{algorithm}
\endgroup
}

If optimizations such as loop reordering and tiling are employed in DNN kernels, the output feature maps can be written to DRAM multiple times within a layer, requiring the VN$_\text{F}$ increment multiple times within a layer (e.g., the VN for the output matrix is incremented twice in the tiled \texttt{MatMul} example).
In this case, the DNN kernel can maintain VN$_\text{F}$ in its program state and keeps track of the VN values associated with the feature maps of each layer in the network.
As the VN$_{\text{F}}$ used for writing each feature map is generated by the DNN kernel on the control processor, the DNN kernel can also provide the VN$_{\text{F}}$ for reading feature maps based on its program state.
Once the VN$_{\text{F}}$ is generated for the feature map, the memory protection unit receives the VN$_{\text{F}}$ value from the control processor and encrypts/decrypts the feature map using the specified value.

\begin{figure}[!t]
\centering
\begin{minipage}{\linewidth}
\begin{subfigure}{.55\textwidth}
 \includegraphics[width=\linewidth, trim={0.4cm 0 0.4cm 0}]{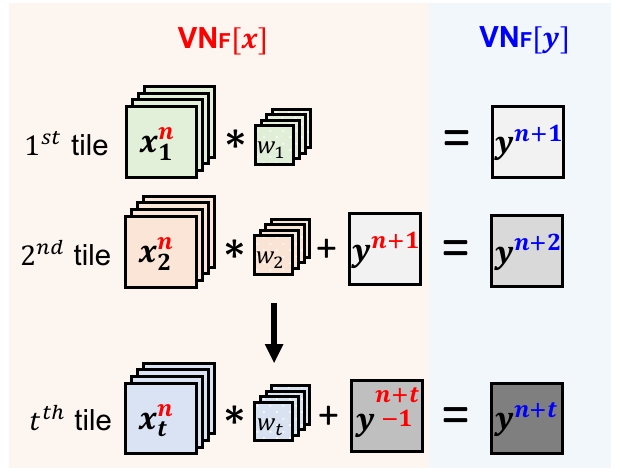}
 \captionof{figure}{VNs for a tiled conv layer --- \small{As the input features $x$ are partitioned into $t$ tiles, the output features $y$ needs to be written to DRAM $t$ times. The control processor sets the VNs for input and output features based on the scheduling.}}
 \label{fig:tiling}
 \end{subfigure}
 \hfill
 \begin{subfigure}{.43\textwidth}
 \dnntiling
 \captionof{figure}{Pseudocode of a conv layer showing how to generate VNs for DRAM accesses --- \small{\texttt{Read/Write} reads and writes the features using the specified VN value}.}
 \label{list:dnn_tiling}
 \end{subfigure}
\end{minipage}%
\caption{Illustration of the VN generation scheme for a tiled convolutional layer.}
\end{figure}

\begin{figure}[!t]
\subfloat[Forward propagation.]{
\begin{minipage}{\linewidth}
    \centering
        \includegraphics[width=3.2in]{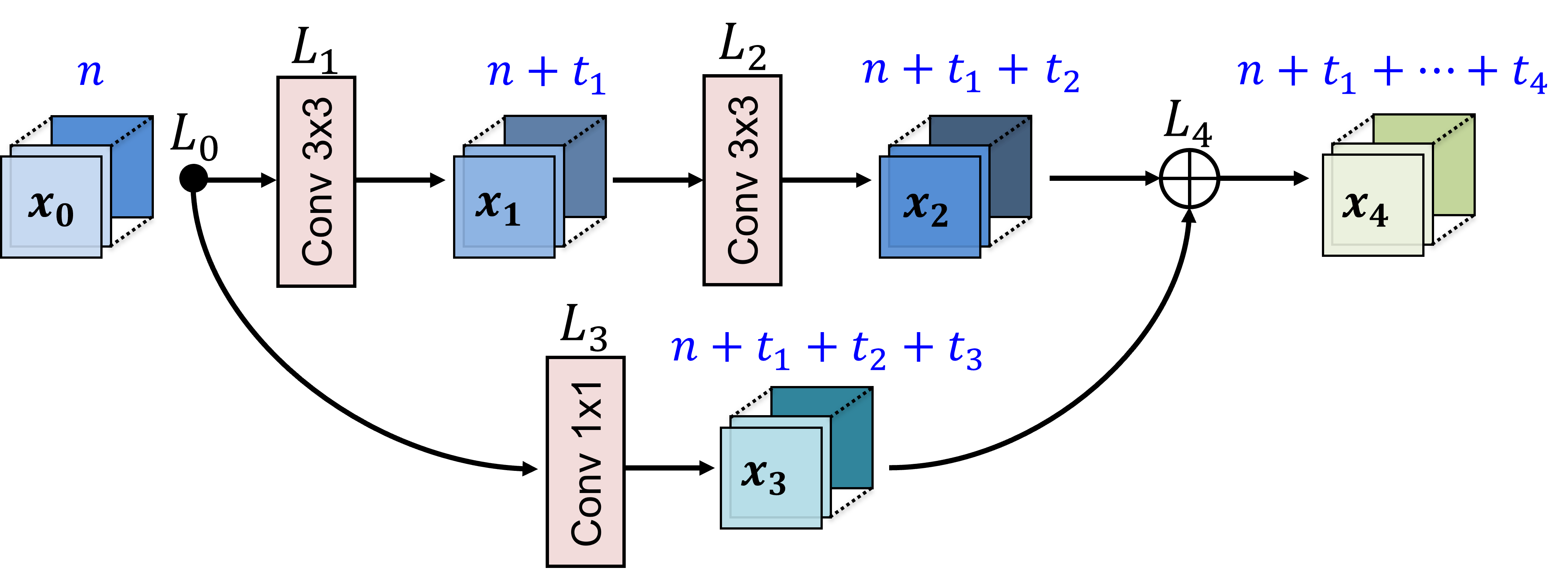}
        \label{fig:residual_forward}
\end{minipage}}
\vspace{-0.05cm}
\subfloat[Backpropagation.]{
\begin{minipage}{\linewidth}
    \centering
        \includegraphics[width=3.2in]{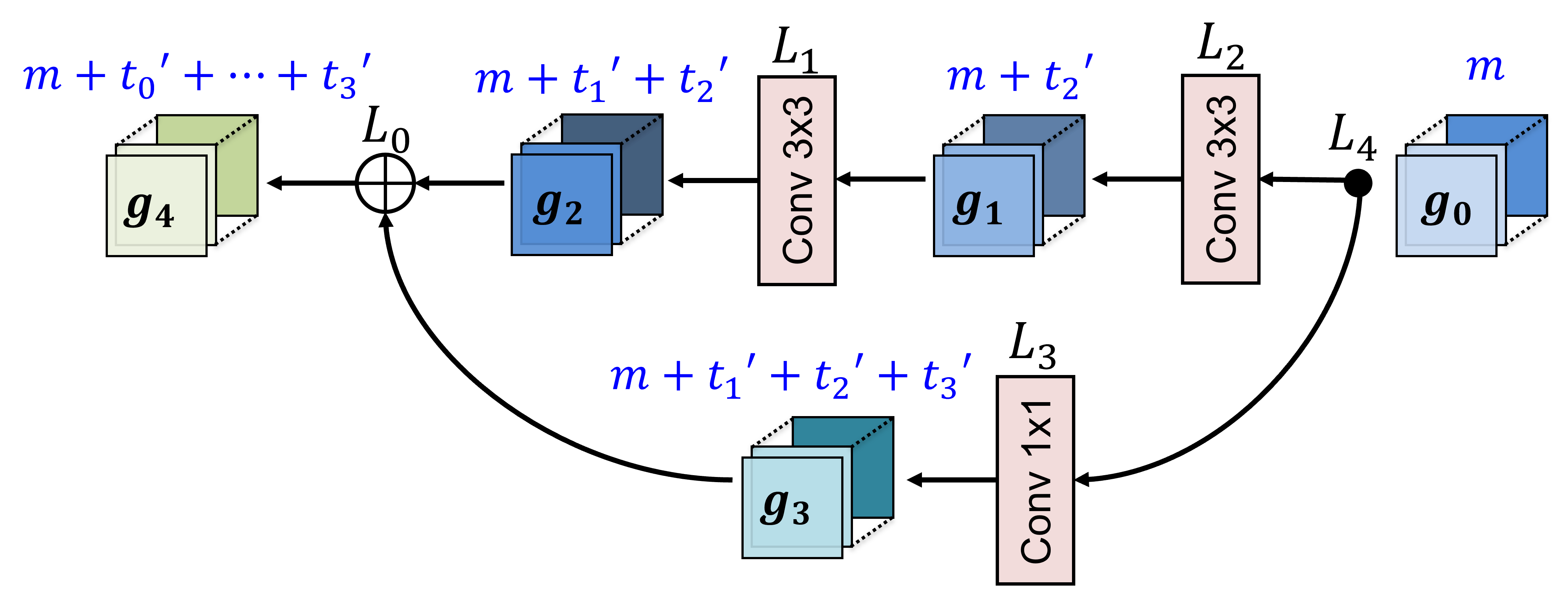}
        \label{fig:residual_backward}
\end{minipage}}
\caption{The VN generation scheme for the forward and backward passes of a residual block --- \small{Blue values represent the VN associated with the feature maps ($x$) and the gradients ($g$). Broadcast and element-wise addition operations in forward pass become the element-wise addition and broadcast operations in backpropagation after differentiation, respectively.}}
\label{fig:residual_block}
\vspace{-0.2cm}
\end{figure}

As illustrated in Figure~\ref{fig:tiling}, the DNN kernel can generate VN$_{\text{F}}$ for the output feature maps $y$, even if $y$ is written to DRAM as many times as the number of tiles.
In this specific case, Algorithm~\ref{list:dnn_tiling} provides the pseudocode of the augmented convolution function running on the control processor for setting the VNs for memory reads and writes.
The VN for the input feature maps $x$ (VN$_{\text{F}}[x]$) remains constant (e.g., $n$) as $x$ is read-only within the layer.
Here, $n$ simply represents the VN for the input feature maps at the beginning of a layer.
In the first iteration, the kernel assigns $n+1$ as the VN value for the output feature maps $y$ (VN$_{\text{F}}[y]$) and writes $y$ with that VN value.
Then, for the rest of the $t-1$ iterations, the kernel reads $y$ with the current VN$_{\text{F}}[y]$ value, increments VN$_{\text{F}}[y]$, and writes the updated $y$ with the new VN$_{\text{F}}[y]$ value.
Assuming that $y$ is written to off-chip memory $t$ times, the final VN value associated with $y$ should be $n+t$.
Similarly, the VNs for feature maps across different DNN layers can also be determined easily by the kernel.
Figure~\ref{fig:residual_forward} shows the computational graph (i.e., DFG) of the forward propagation of a residual block, which is a widely used structure in many modern DNNs~\cite{he2015resnet, howard2017mobilenets}.
Suppose each conv layer ($L_1$, $L_2$, $L_3$) and the element-wise addition layer ($L_4$) write their output to DRAM $t_1$, $t_2$, $t_3$, and $t_4$ times, respectively.
The DNN kernel can compute the VN for each feature map in the residual block based on the scheduling as $\text{VN}_\text{F}[x_i] = n + \sum_{k=1}^{i} t_k$, where $n$ is the VN for the input features to the residual block.
The VN value of the output features is incremented for different layers to guarantee that the same counter value is never reused, even though output features from different layers can exist in memory at the same time.

The weights are read-only during inference.
Therefore, we can use a constant as the VN for the weights until they are updated.
To allow updating weights, the DNN kernel tracks VN$_{\text{W}}$ in its program state and
keeps track of the number of updates (writes) to the weights.

Note that VN$_\text{F}$ and VN$_\text{W}$ are all kept as part of the program state in the trusted control processor,
and there is no VN stored in external memory.
For simplicity, the kernel can maintain one VN$_\text{F}$ for the output features of each layer and one VN$_\text{W}$ for the entire network.
In this implementation, a 127-layer DNN uses 1 KB on-chip state for VNs.
The size of the on-chip state can be reduced by only tracking the non-consumed features (i.e., the features will be used as the input to later layers) or leveraging the network structure statically known to the kernel. 
For example, the VN for output features can be determined from one counter that keeps the number of inputs processed and the layer number.
For the 64-bit VN in our design, an accelerator with a throughput of
1,000 inputs per second for a 1,000-layer DNN can run for 0.28 million years before an overflow.
If an overflow happens, MGX requires the memory to be re-encrypted with a new key.

\noindent\textbf{DNN Training --}
One iteration of training consists of a forward propagation and a backpropagation. 
The forward propagation is the same as inference except that all intermediate features 
are saved, %
and can use the VN generation strategy for inference. 
Here, we describe the VN generation for the backpropagation. 
Each layer first computes the gradients flowing to the previous layer using the gradients flowing into current layer and the associated weights.
Then, the layer's weights are updated using the incoming gradients and the saved features.

VNs are constructed in the same way as in the inference.
The backpropagation only adds additional feature reads and
does not affect the VN generation for features.
The VNs for weights still use VN$_\text{W}$ as all weights are updated the same number of times.
However, VN$_\text{W}$ are incremented more frequently, where VN$_\text{W}$ tracks the number of updates to the weights, where the weights are updated exactly once during backpropagation.
Gradients in backward pass correspond to the feature maps in forward pass.
As illustrated in Figure~\ref{fig:residual_backward}, the VNs for gradients can be computed in a way similar to computing the VNs for feature maps.
In addition to VN$_\text{F}$ and VN$_\text{W}$, the control processor also needs to keep track of the VN of the gradients (VN$_\text{G}$) associated with each layer.
It is worth noting that broadcast ($L_0$) and element-wise addition ($L_4$) operations in forward pass become the element-wise addition and broadcast operations in backpropagation, respectively. Assuming that the $i^{th}$ layer writes its output to DRAM $t_i'$ times, the VN for each gradient tensor can be written as shown in Figure~\ref{fig:residual_backward}, where $m$ is the VN value for the input gradients ($g_0$) of the residual block. 

\section{MGX for Graph Processing}
This section provides background on graph processing and discusses how MGX can be applied to graph accelerators for different graph algorithms, such as PageRank and Breadth-First Search (BFS).

\subsection{Background on Graph Algorithms}
Graphs represent a popular way to encode connections in many important applications including social networks, electrical grid, circuits, etc. 
At the same time, processing large graphs requires high performance.  
To achieve high performance and low power, 
many ASIC/FPGA accelerators are proposed for graph processing.

In this work, we focus on the GraphBLAS formulation~\cite{graph_blas}, where key graph processing operations (e.g., traversals, shortest path) are formulated as sparse linear algebra operations such as sparse-matrix dense-vector multiplications (SpMV). 
GraphBLAS extends the expressiveness of linear algebra in representing graph operations by leveraging the concept of semiring. 
A semiring is defined as a 5-tuple ($\mathbb{D}, \otimes, \oplus, I_\otimes, I_\oplus$), where $\mathbb{D}$ is a set, $\otimes$ is a scalar multiplication operation, $\oplus$ is a scalar addition operation, $I_\otimes$ is the identity of $\otimes$, and $I_\oplus$ is the identity of $\oplus$.
The matrix operations on a semiring can correspond to a step in many different graph algorithms, depending on specific operators used in that semiring. 
For example, PageRank, BFS, and single-source shortest path (SSSP) can be expressed using the following semirings:
\begin{itemize}
    \item PageRank: ($\mathbb{R}$, $\times$, $+$, 1, 0)
    \item BFS: (Boolean, $\&$, $\lvert$, 1, 0)
    \item SSSP: ($\mathbb{R} \cup \infty$, $+$, min, 0, $\infty$)
\end{itemize}
Given the expressiveness and the flexibility of GraphBLAS, an ASIC/FPGA accelerator designed with the GraphBLAS programming interface can be used to execute a rich set of graph algorithms, whereas many existing graph processing accelerators are designed to accelerate one specific graph algorithm such as PageRank and BFS~\cite{pagerank_fpga, bfs_fpga}. 
Therefore, we investigate the applicability of MGX with a focus on GraphBLAS-based graph processing accelerators.

GraphBLAS represents the topology of a graph (i.e., the connectivity between vertices) as a sparse adjacency matrix and the attributes associated with vertices as a sparse or dense vector.
\begin{figure}[t!]
\centering
\includegraphics[scale=0.35]{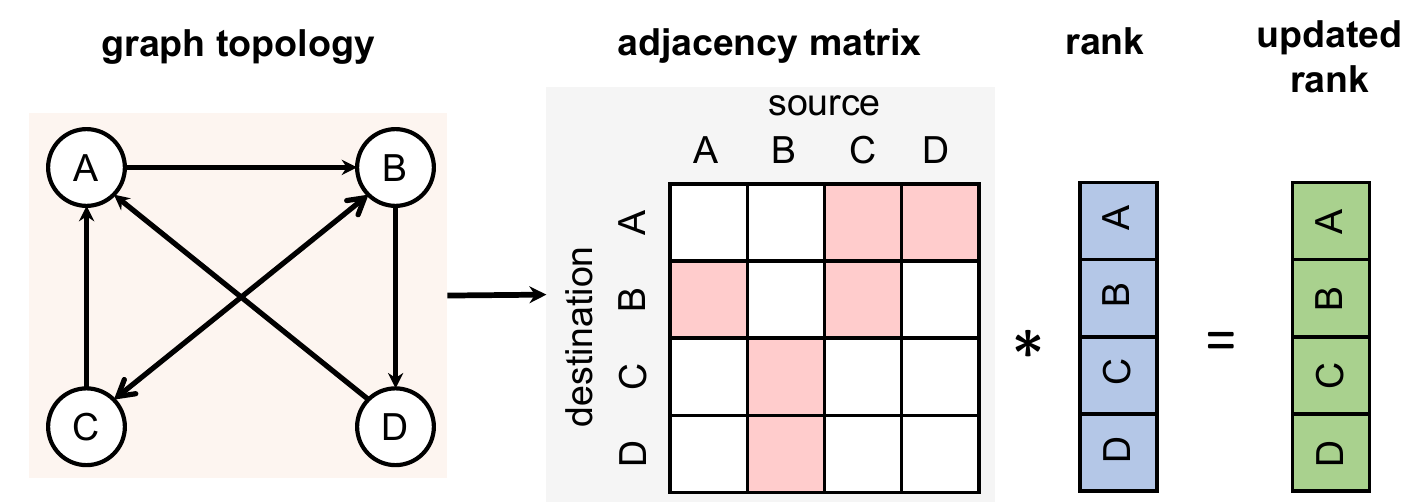}
\caption{Example graph of PageRank algorithm --- \small{The sparse adjacency matrix encodes the graph topology, where each cell represents the weight between the two connected vertices. The rank vector represents the current value of the attributes associated with each vertex whereas the updated rank vector holds the attribute values for the next iteration}.}
\label{fig:example-graph}
\end{figure}
Figure~\ref{fig:example-graph} provides an example for the presentation of the graph topology and attribute values (i.e., the ranks of vertices) in PageRank using GraphBLAS.
The graph topology is encoded with the adjacency matrix, where each non-empty cell represents the weight between the two connected vertices.
The rank vector contains the current attribute values associated with the vertices.  
Then, PageRank algorithm can be expressed as an SpMV operation between the sparse adjacency matrix and the dense rank vector.
The resulting updated rank vector remains in a dense format and will be used as the rank vector for the next iteration.

\subsection{Version Number Generation for GraphBLAS}
MGX is also applicable to graph processing accelerators.
Here we discuss how to apply MGX to a GraphBLAS-based accelerator using the PageRank algorithm as an example. 
While our discussion mainly focuses on PageRank, MGX can also be applied to other graph algorithms supported by GraphBLAS.

PageRank is an iterative algorithm, which computes the rank of each vertex by calculating the likelihood of that vertex being reached.
There are three main data structures to store a graph in PageRank --- a sparse adjacency matrix, a dense rank vector, and a dense updated rank vector.
The sparse adjacency matrix stores all edges in a graph, which is represented as a tuple of the IDs of the source and destination vertices.
Due to the sparsity of the graph, the adjacency matrix is usually stored in a compressed sparse format to eliminate redundant memory accesses.
Both rank and updated rank vectors are stored in dense format.
Each entry in the three data structures typically occupies 4 bytes in memory and each data structure can have several to thousands of millions of entries in real-world graphs~\cite{graph-dataset1, graph-dataset2, hu2020ogb}. 

\begin{figure}[t!]
\centering
\includegraphics[scale=0.38]{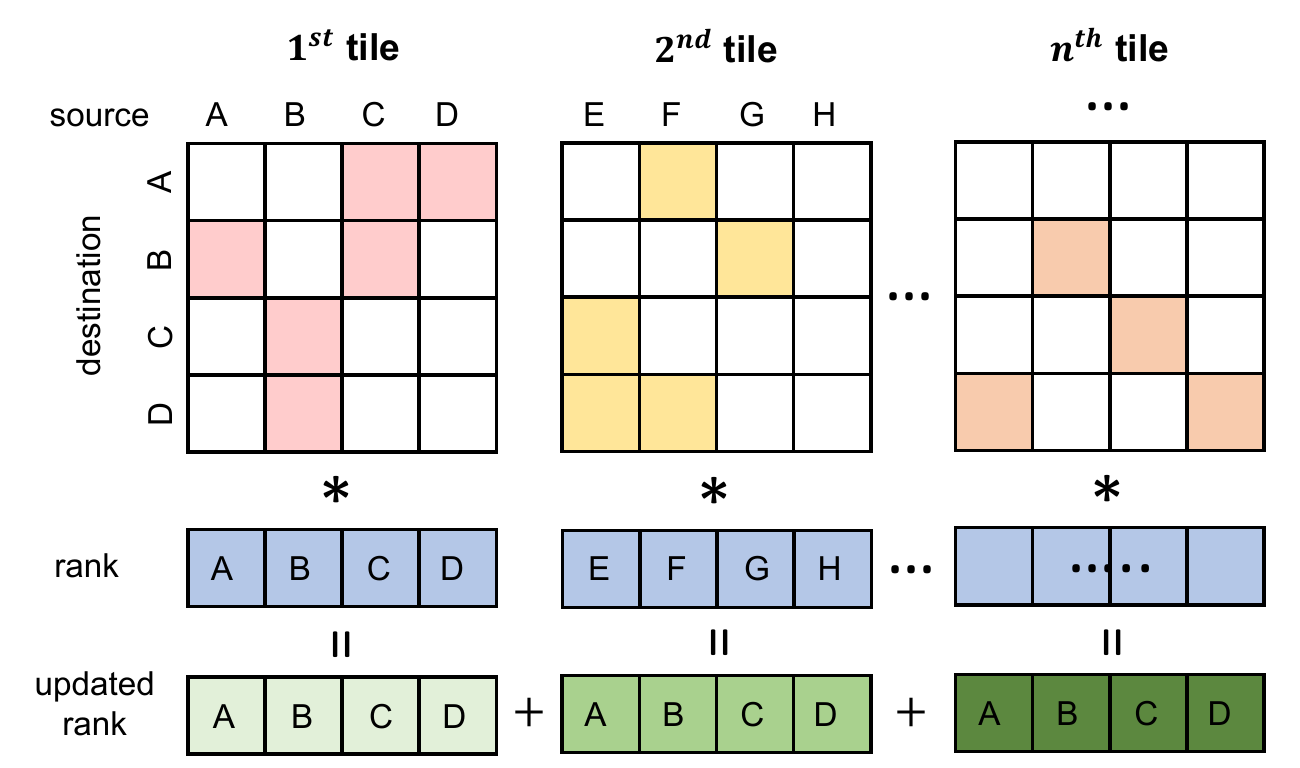}
\caption{The scheduling of a graph accelerator.}
\label{fig:graph_scheduling}
\end{figure}

In each iteration of PageRank, the graph processing accelerator needs to update the attribute values of all vertices (i.e., calculate the updated rank vector).
GraphBLAS-based accelerators typically use the scheduling as depicted in Figure~\ref{fig:graph_scheduling}.
The graph processing accelerator computes the updated rank vector for a subset of vertices (e.g., $\{A, B, C, D\}$) at a time and generates the updated rank vector for all vertices in a sequential manner. 
When calculating the updated rank for a subset of destination vertices (e.g., $\{A, B, C, D\}$), the accelerator accesses a tile of the sparse adjacency matrix between these destination vertices and a subset of source vertices (e.g., $2^{nd}$ tile) and the corresponding rank vector (e.g., the rank $\{E, F, G, H\}$) to get the partial result of the updated rank vector.
After processing all tiles, the final results of the updated ranks for the subset of destination vertices are obtained.

In PagePank, the sparse adjacency matrix is read-only and read sequentially as the adjacency matrix is stored in a sparse format.
In addition, the adjacency matrix remains unchanged across different iterations of PageRank.
Thus, MGX can assign a constant VN value for the adjacency matrix.
However, the size of the sparse adjacency matrix in each tile differs because each destination vertex may be connected to an arbitrary set of source vertices.
For example, in Figure~\ref{fig:graph_scheduling}, the first, second, and third tiles contains six,  five, and four edges, respectively.
Because the sparse adjacency matrix needs to be read with an irregular block size, integrity verification for the adjacency matrix must be in a fine granularity to avoid unnecessary memory reads.
For example, an accelerator may read/write 64-byte chunk of the adjacency matrix at a time, and have a MAC for each chunk.

Instead, MGX can calculate the coarse-grained MAC exploiting the fixed scheduling of the graph processing accelerator.
As the sparse adjacency matrix remains unchanged, the accelerator partitions the sparse adjacency matrix in the same way across different iterations.
Therefore, MGX can protect the confidentiality and integrity of the sparse adjacency matrix at the tile level, where all elements in the same tile can be protected using a single MAC.
The rank vector is also read-only and read sequentially during each iteration, only requiring one VN per graph.
Each tile of the updated rank vector is written sequentially and will be written to the off-chip memory for the same number of times, which also only requires one VN per graph.
Therefore, the kernel on the control processor only needs to track the number of executed iterations (Iter) of PageRank to calculate VNs, thus requiring only 64-bit additional on-chip state.
Specifically, $(\text{Iter}-1)$ is used as the VN when reading a tile of the rank vector and Iter is used as the VN when writing a tile of the updated rank vector.
With MGX, the VN can be computed without off-chip memory accesses, eliminating the overhead of memory encryption.
MACs can be calculated at coarse granularity to reduce the overhead of integrity verification.
Since BFS uses the same SpMV operation as PageRank, the VN generation scheme remains the same and only one Iter counter is added to the accelerator state.

In addition to the SpMV operation adopted in PageRank, sparse-matrix sparse-vector multiplication (SpMSpV) is another important linear algebra operation in GraphBLAS.
Compared with SpMV, the only difference is that the SpMSpV operation reads the attribute values associated with the vertices randomly instead of sequentially.
MGX can still use the same VN generation scheme for all data structures and the same MAC granularity for the adjacency matrix and the vector with updated attribute values, as in SpMV.
However, the vector holding the current attribute values requires a fine-grained MAC.
In this case, MGX can still greatly reduce the overhead of off-chip memory protection.
\section{Evaluation}

\subsection{Accelerator and Simulation Setup}

For DNN acceleration, we use cycle-level simulations to (1) compare the performance overhead of multiple memory protection schemes,
(2) study the overhead for a larger class of DNN models, and (3) evaluate the overhead for DNN inference and training.
Specifically, we use SCALE-Sim~\cite{scale-sim1}, an open-source cycle-level DNN accelerator simulator from ARM research.
For graph processing, we use a combination of RTL and cycle-level simulation to compare the performance overhead of different memory protection schemes.
In particular, we use GraphLily~\cite{graphlily}, an open-source GraphBLAS accelerator written in HLS.

\begin{figure}[t]
\begin{center}
\subfloat[DNN accelerator.]{
\begin{minipage}{.47\linewidth}
    \centering
        \includegraphics[width=\columnwidth]{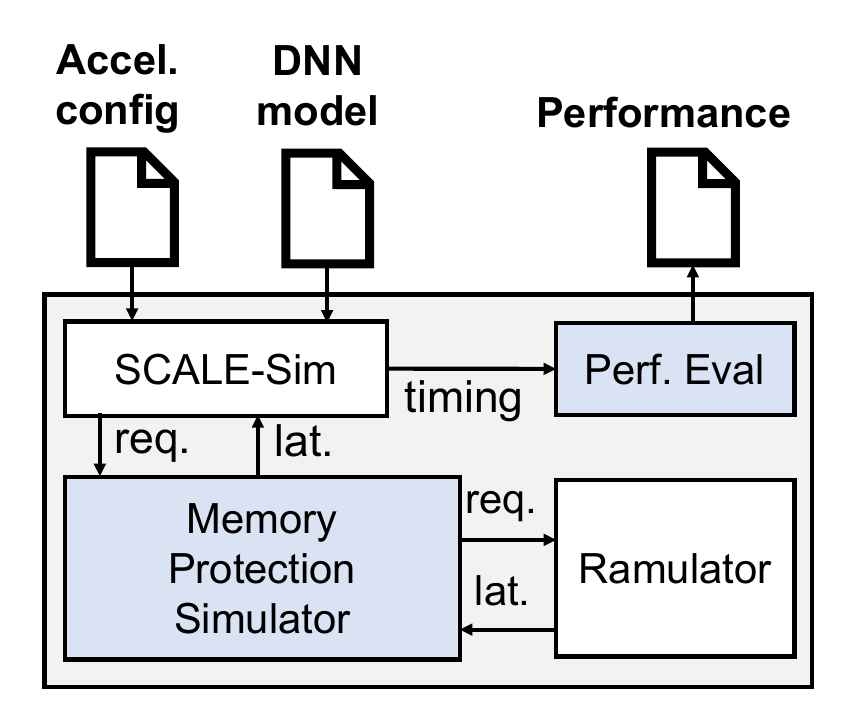}
        \vspace{-0.5cm}
        \label{fig:dnn_sim}
\end{minipage}}
\hspace{0.04cm}
\subfloat[Graph processing accelerator.]{
\begin{minipage}{.47\linewidth}
    \centering 
        \includegraphics[width=\columnwidth]{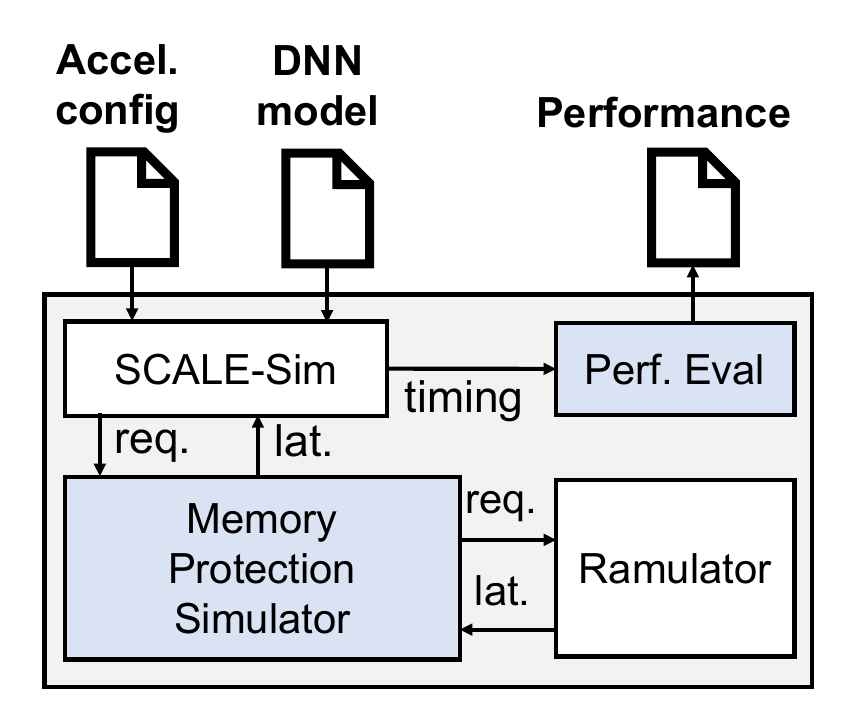}
        \vspace{-0.5cm}
        \label{fig:graph_sim}
\end{minipage}}
\end{center}
\caption{The block diagram of the cycle-level simulator for the secure accelerators.}
\vspace{-0.25cm}
\label{fig:flow}
\end{figure}

As shown in Figure~\ref{fig:flow}, both simulation setups have three main components --- an accelerator, a memory protection simulator, and off-chip memory.
The DNN accelerators are simulated in a cycle-level DNN simulator (i.e., SCALE-sim) to generate a trace of computation and memory events.
The HLS graph processing accelerator (i.e., GraphLily) is first synthesized into RTL design.
We then use RTL simulation to obtain the trace of computation and memory events.
After the memory traces are obtained, a memory protection simulator uses the event trace to calculate the total execution time and the bandwidth usage by simulating protection mechanisms and DRAM accesses.
The memory accesses are simulated using Ramulator~\cite{ramulator} DDR4 at 2400MHz.
The performance of the accelerator is maximized when the memory bandwidth matches the computation throughput, which means the accelerator is neither compute nor memory bounded.
Finally, a performance evaluator generates the final performance based on the timing of computation and new memory events (including additional memory events from memory protection).

\noindent\textbf{Accelerator Configurations --}
To evaluate the MGX for a DNN accelerator under different use cases, we model a large and a small configurations, namely \texttt{Cloud} and \texttt{Edge}, for cloud and edge computing, respectively.
\texttt{Cloud} is modeled based on Google TPU-v1~\cite{google2017tpu} and \texttt{Edge} uses a similar configuration as the Samsung Neural Processing Unit~\cite{samsung-npu}.
Specifically, \texttt{Cloud} and \texttt{Edge} contain 64k and 1k processing elements (i.e., MAC units) and 24 MB and 4.5 MB on-chip memory, running at 700 MHz and 900 MHz, respectively.
To balance computation and memory bandwidth, we simulate one 64-bit DDR channel for \texttt{Edge} and four 64-bit DDR channels for \texttt{Cloud}.
The size of the protected memory is 16~GB.
For the graph accelerator, we simply adopt the original design of the GraphLily accelerator.
The clock frequency of the graph accelerator is assumed to be 800 MHz.
We simulate four 64-bit DDR channels at 2400 MHz to provide enough bandwidth.

\noindent\textbf{Benchmarks --}
For the DNN accelerator, we evaluate MGX on a variety of DNN architectures --- AlexNet, VGG, GoogleNet, and ResNet for image classification and BERT (i.e., Transformer encoder) for language pretraining, and DLRM for personalized recommendation.
For each DNN model, we simulate both inference (forward propagation) and training (forward propagation and backpropagation) of the network.
The weight update during training is not emulated as no similar operation is available in SCALE-Sim.

For the graph accelerator, we validate the effectiveness of MGX on two widely-used graph algorithms: PageRank and BFS. Both algorithms are executed using the SpMV engine in GraphLily.
We perform PageRank and BFS on six existing graph benchmarks, including Google-plus, pokec, livejournal, and reddit from the Stanford Network Analysis Project~\cite{snapnets} and ogbl-ppa and obgn-products from the Open Graph Benchmark~\cite{hu2020ogb}.
Google-plus, pokec and livejournal are social network graphs. 
reddit is composed of posts from the Reddit forum.
ogbl-ppa and obgn-products are two large graph datasets containing 576K and 2449K vertices and 42M and 124M edges, respectively.

\noindent\textbf{Memory Protection --}
We implement the recent memory encryption engine (MEE) design from Intel~\cite{cpu_MEE} as the baseline memory encryption.
This baseline uses a multi-level 8-ary Merkle tree with 56-bit \VNs~and MACs, and works at
a 64-byte granularity.
Similarly, for integrity verification, we implemented the baseline that uses one MAC for each
64-byte block.
Because the DNN accelerator has a largely streaming memory access pattern, increasing the \VN/MAC cache does not help unless it is big enough to capture temporal locality across layers.
In our experiments, we include a reasonably large (32-KB) on-chip cache for \VNs~and MACs in the baseline scheme.
The \VN/MAC cache uses the LRU replacement policy with write-back and write-allocate policies.
MGX has no cache for \VNs~and MACs, and protects the integrity using a MAC per 512-byte block for most applications, amortizing the overhead of memory protection over a large chunk of data.
It is worth noting that the MAC granularity of the embedded tables in DLRM is still 64-byte, as fine-grained access to the embedded tables is required.

\subsection{Experimental Results}
\noindent\textbf{Performance --}
We compare the accelerator performance for
three different protection schemes: no protection (NP), 
today's baseline memory protection (BP), and MGX.
The results for BP and MGX are normalized to the one with no protection.

\begin{figure}[t]
\subfloat[Inference.]{
\begin{minipage}{\linewidth}
    \centering
        \includegraphics[width=3.3in]{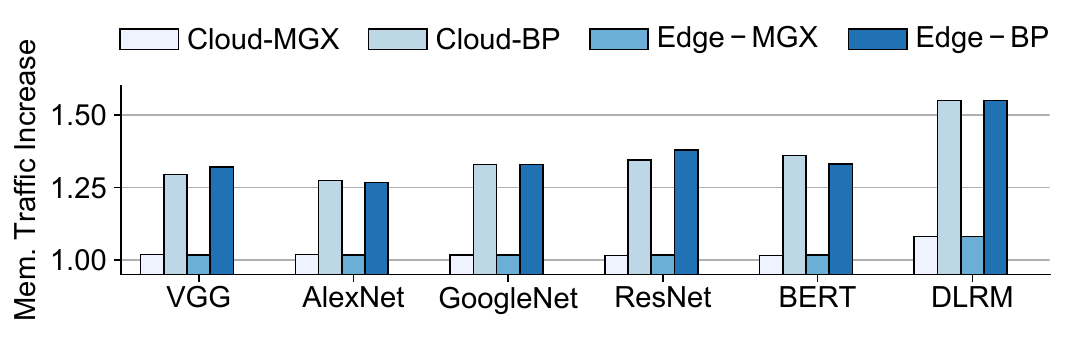}
        \vspace{-0.1cm}
        \label{fig:bw_inf}
\end{minipage}}\hfill
\subfloat[Training.]{
\begin{minipage}{\linewidth}
    \centering
        \includegraphics[width=3.3in]{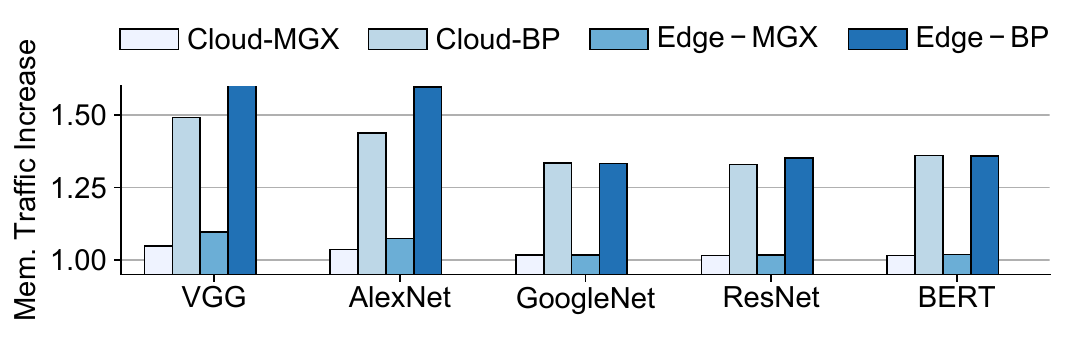}
        \vspace{-0.1cm}
        \label{fig:bw_train}
\end{minipage}}
\caption{The memory traffic increase of DNN inference and training.}%
\label{fig:mem_traffic}
\vspace{-0.2cm}
\end{figure}

Figure~\ref{fig:mem_traffic} compares the memory traffic increase of two DNN accelerator configurations with MGX and BP --- Cloud-MGX, Cloud-BP, Edge-MGX, Edge-BP.
Cloud-BP and Edge-BP introduce 36.0\% and 36.3\% more memory accesses on average for inference, respectively.
In particular, the inference of the recommendation model (i.e., DLRM) increases the memory traffic by 55\%.
For training, the average increases in memory accesses are 37.8\% and 42.9\% for Cloud-BP and Edge-BP.
The memory traffic increase is larger for training because the training process accesses more data and has more frequent cache evictions in the \VN/MAC cache.
Cloud-MGX and Edge-MGX increase the memory traffic by an average of \textbf{2.4\%} and \textbf{2.4\%} for inference and \textbf{2.7\%} and \textbf{3.5\%} for training, respectively.
The results demonstrate the advantage of the MGX, which removes VNs 
stored in DRAM and uses a MAC per 512-byte data block to 
match the accelerator's data movement granularity.

\begin{figure}[t]
\subfloat[Inference.]{
\begin{minipage}{\linewidth}
    \centering
        \includegraphics[width=3.3in]{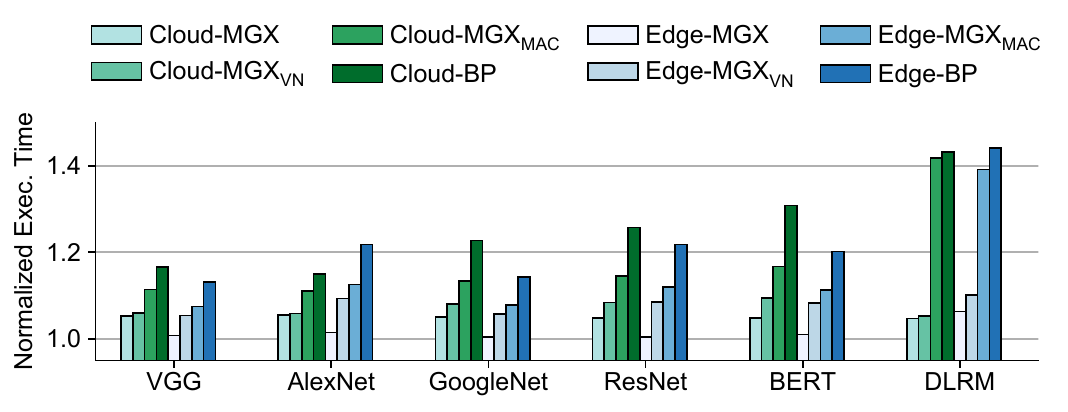}
        \vspace{-0.1cm}
        \label{fig:lat_inf}
\end{minipage}}\hfill
\subfloat[Training.]{
\begin{minipage}{\linewidth}
    \centering
        \includegraphics[width=3.3in]{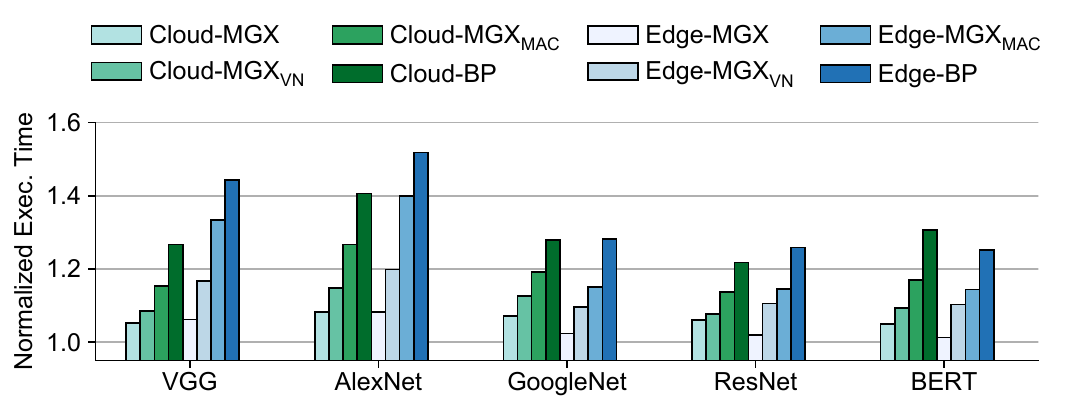}
        \vspace{-0.1cm}
        \label{fig:lat_train}
\end{minipage}}
\caption{The normalized execution time of the DNN inference and training on different networks models.}
\label{fig:lat}
\end{figure}

Figure~\ref{fig:lat} shows the performance of the baseline protection and MGX.
Cloud-BP and Edge-BP are 1.24$\times$ and 1.32$\times$ slower than no protection on average for inference and training.
For the cloud and edge accelerators, MGX achieves a much smaller performance overhead than BP; the average overhead is \textbf{3.2\%} for inference and \textbf{4.7\%} for training.
To better understand the contribution of each optimization to the overhead reduction, we include the results of two MGX variants, MGX$_{\text{VN}}$ and MGX$_{\text{MAC}}$, which use only one optimization: on-chip VN generation or coarse-grained MAC.
MGX$_{\text{VN}}$ is 1.08$\times$ and 1.12$\times$ slower than no protection on average for inference and training.
MGX$_{\text{MAC}}$ has a higher overhead than MGX$_{\text{VN}}$, on average 1.16$\times$ and 1.20$\times$ slower than no protection for inference and training.
This result shows that both on-chip VN generation and coarse-grained MAC are important in reducing the overhead of off-chip memory protection.

\begin{figure}[t]
\subfloat[Memory traffic increase.]{
\begin{minipage}{\linewidth}
    \centering
        \includegraphics[width=3.3in]{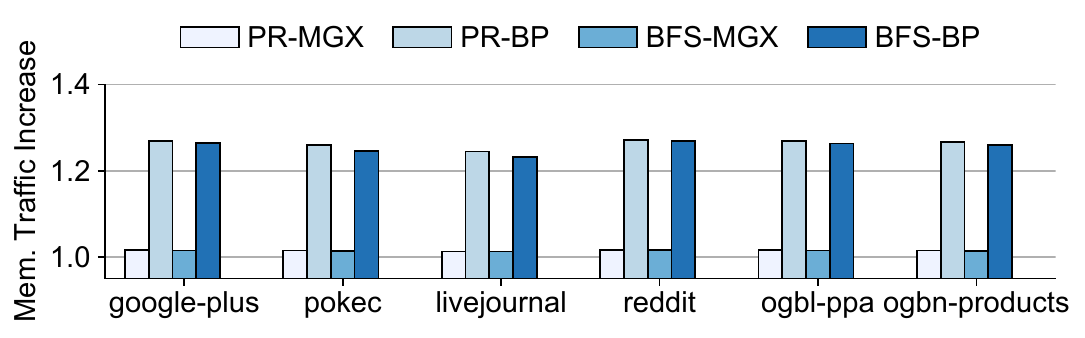}
        \vspace{-0.1cm}
        \label{fig:bw_graph}
\end{minipage}}\hfill
\subfloat[Normalized execution time.]{
\begin{minipage}{\linewidth}
    \centering
        \includegraphics[width=3.3in]{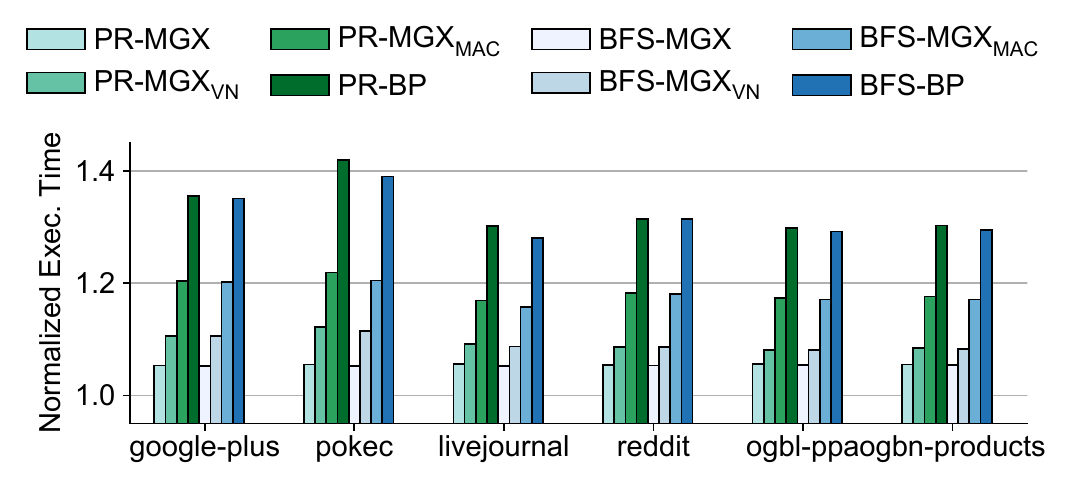}
        \vspace{-0.1cm}
        \label{fig:lat_graph}
\end{minipage}}
\caption{The memory traffic increase and the normalized execution time of PageRank (PR) and BFS.}
\label{fig:graph_results}
\end{figure}

For the graph accelerator, we compare the memory traffic increase and execution time of PageRank and BFS with MGX and BP --- PageRank-MGX, PageRank-BP, BFS-MGX, and BFS-BP.
As shown in Figure~\ref{fig:bw_graph}, PageRank-BP and BFS-BP introduce 26.3\% and 25.6\% more memory accesses on average, respectively.
MGX only adds \textbf{1.5\%} and \textbf{1.4\%} additional memory accesses for PageRank and BFS, respectively.
Compared to BP, MGX is able to significantly reduce the meta-data memory accesses, demonstrating the effectiveness of VN generation and coarse-grained MAC.

Figure~\ref{fig:lat_graph} compares the performance of the baseline protection and MGX for PageRank and BFS.
BP leads to a significant slowdown for both PageRank and BFS.
For PageRank and BFS, the maximum slowdown due to BP is 1.42$\times$ and 1.39$\times$, respectively.
In contrast, MGX introduces only the maximum overhead of \textbf{5.2\%} for both graph algorithms.
Across all benchmarks, BP and MGX have average performance overhead of 32.7\% and \textbf{5.0\%}.
In addition, the average performance overheads of MGX$_{\text{VN}}$ and MGX$_{\text{MAC}}$ are 9.4\% and 18.0\% across all benchmarks.

\subsection{Case Study on Existing Accelerators}
To show how MGX can be applied to existing DNN accelerators, we study CHaiDNN~\cite{chaidnn}, which is an open-source DNN accelerator from Xilinx.
CHaiDNN has a relatively simple accelerator interface, which only supports three high-level operations including Convolution, Deconvolution, and Pooling.
Activation functions are merged with high-level operations to avoid unnecessary DRAM access and to maximize performance.
Because of the high abstraction level of CHaiDNN, a deep neural network like AlexNet can be expressed in less than 20 instructions.

In order to equip CHaiDNN accelerator with MGX, we can implement the MGX scheme using a microcontroller for generating and managing version numbers.
Each CHaiDNN instruction can be treated as a DNN layer.
For each layer, the microcontroller assigns a VN value to all output features belonging to that layer and keeps track of the VN values in the on-chip VN table (i.e., the microcontroller's SRAM memory).
The VN table also needs to have two counters for weights and inputs as described in Section~\ref{sec:dnn_vn}.
In addition to the microcontroller, we also need to add AES Galois/Counter Mode (AES-GCM) cores~\cite{aes_gcm} for both memory encryption and integrity verification.
As the DNN accelerators typically have large processing element arrays and on-chip buffer, the overhead of adding microcontroller and AES-GCM cores is expected to be modest.

\section{Discussion}
In this section, we discuss two additional cases to show MGX is also applicable to other accelerators such as genome alignment and video decoding. 
In addition, we also show that MGX can be applied to DNNs with static and even dynamic pruning techniques.

\subsection{Applicability of MGX}
\label{sec:applicability}

\begin{figure}
    \begin{minipage}{\columnwidth}
	\centering
	\includegraphics[width=\columnwidth]{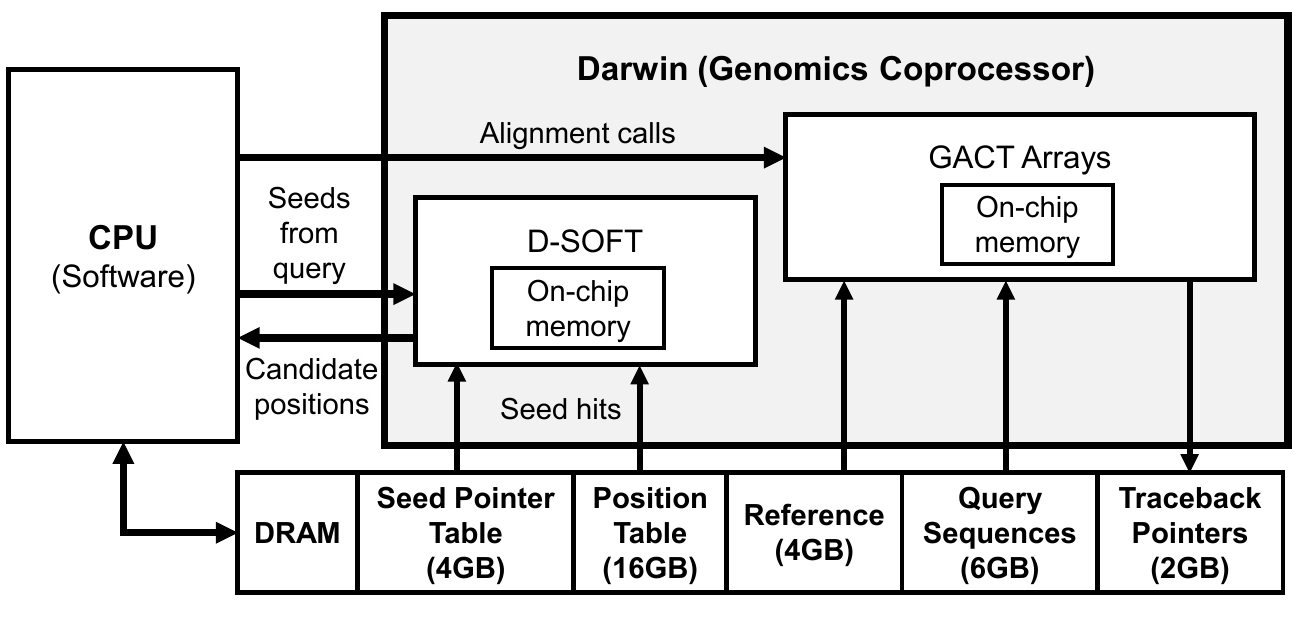}
	\captionof{figure}{Block diagram of Darwin accelerator.}
	\vspace{0.1cm}
	\label{fig:darwin-block}
	\end{minipage}
\end{figure}

\noindent\textbf{Genome Alignment --} 
We consider the off-chip memory protection for  
Darwin~\cite{darwin}, which is an accelerator for genome assembly.

While Darwin also relies on a CPU to perform certain initialization operations and control
the hardware acceleration, we assume that the CPU and its communications with Darwin are protected 
separately with a secure computing technology (e.g., Intel SGX) and 
focus on protecting memory accesses for the accelerators in this discussion.

Figure \ref{fig:darwin-block} shows the components and data accesses in Darwin.
Darwin consists of two hardware-accelerated parts, D-SOFT and GACT, which use 
five types of data in off-chip memory: reference sequences, a seed-pointer table, 
a position table, query sequences and traceback pointers.
During initialization for a reference-assisted assembly, the reference sequence, the seed-pointer table, and the position table
are loaded (written) into memory once by a CPU; these are later only read by the accelerator.
Therefore, the version number for these three data structures can be obtained simply from a counter in the on-chip state of the accelerator, 
which increments on each new genome assembly (CTR$_\text{genome}$).

  \begin{figure}[t]
  		\begin{minipage}{\linewidth}
  			\centering
  			\includegraphics[width=3.3in]{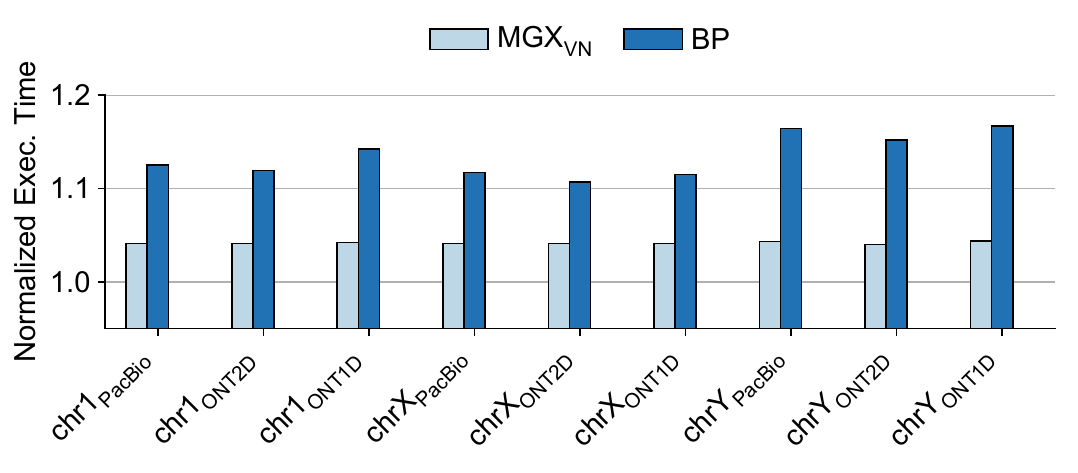}
  			\vspace{-0.1cm}
  	\end{minipage}
  	\caption{The normalized execution time of various GACT workloads.}
\label{fig:Genome_perf}
\end{figure}
  
After initialization, the CPU loads a batch of query sequences into memory and runs D-SOFT and GACT on 
the accelerator for each query in the batch. 
D-SOFT generates a filtered list of candidate positions from seeds in the query that hit in the reference sequence.
These are passed on to GACT arrays as tiles for extension i.e. alignment.
During these processes, the seed pointer \& position tables, and the query \& reference sequences are all only read by the accelerator.
The output consists of GACT arrays writing traceback pointers for each tile sequentially into the memory.
Hence, for the query sequences and traceback pointers, we can keep a counter in the accelerator state that increments for each new
query batch (CTR$_\text{query}$), and use the concatenation of CTR$_\text{genome}$ and CTR$_\text{query}$ as the version number. 
The traceback pointers are later processed by the software to construct aligned reads.

The GACT part of Darwin is available as open-source RTL, whereas D-SOFT is available as software. 
We evaluate the performance of GACT for reference-guided assembly,
using the latest human genome assembly GRCh38 \cite{grch38} as reference. For chromosomes 1, X \& Y, we generate three sets of reads simulating different sequencers (PacBio, ONT2D, and ONT1D)  with varying error profiles, as described in \cite{darwin}.

We use the D-SOFT software to generate a list of candidate positions (or tiles) that are sent to the GACT hardware for alignment.
For each tile, a GACT array loads a chunk of reference and query sequences from a specified DRAM offset, performs alignment of that tile, and finally writes the traceback pointers to DRAM.
For the memory accesses of each tile, we obtain the data transfer times through a memory protection simulator, 
with four DDR4-2400 channels.
For obtaining the computation time of the tiles, we perform an RTL simulation of GACT, 
with the same settings as specified in \cite{darwin}.
We assume an ASIC configuration with 64 GACT arrays that can process tiles independently, each containing 64 PEs, running at 800 MHz. 
As D-SOFT generates calls to GACT for  millions of tiles, we simulated only a subset of the tiles.
The memory and computation times are used to calculate the overall execution time.
Because GACT loads input chunks from effectively random locations in the reference and non-contiguous locations in the query, and since the tile size can be variable,
we do not use coarse-grained MACs for GACT, and only simulate the MGX$_{\text{VN}}$ mode with on-chip VN generation and fine-grained MACs.

We first compare the memory traffic increase of the baseline protection scheme (BP) with MGX$_{\text{VN}}$.
The elimination of off-chip VNs leads to a reduction in memory traffic overhead from 34\% in BP to \textbf{12.5\%} in MGX$_{\text{VN}}$,
the remaining overhead coming from the fine-grained MACs.
Figure \ref{fig:Genome_perf} shows the performance overhead of GACT.
The average performance overhead for BP is $14\%$.
The performance overhead of genome assembly alignment is lower than DNN and graph algorithms because the Darwin accelerator design is more compute-bound.
MGX$_{\text{VN}}$ further reduces the average performance overhead to only \textbf{4\%}.

\begin{figure}[t]
	\centering
	\includegraphics[width=\columnwidth]{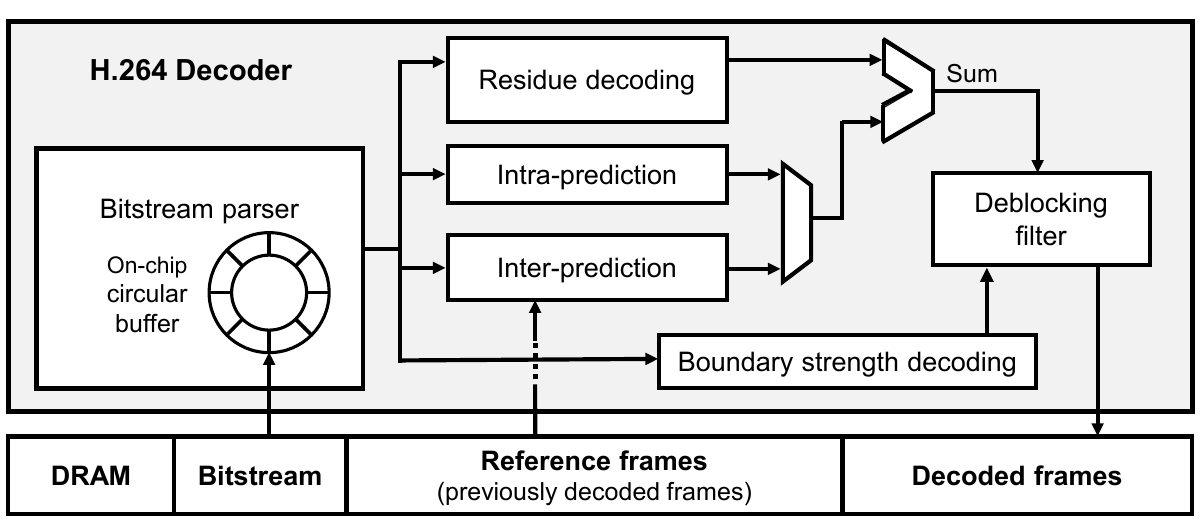}
	\caption{Block diagram of a typical H.264 decoder --- \small{the Inter-prediction module reads reference frames from the off-chip memory for constructing predicted frames}.}
	\vspace{0.2cm}
	\label{fig:h264-block}
\end{figure}

\noindent\textbf{H.264 Video Decoding --} 
We studied H.264/AVC video decoding \cite{iso_h264avc} as another candidate for MGX memory protection.
Figure \ref{fig:h264-block} shows a typical H.264 decoder architecture, which transforms an input bitstream into video frames. 
The input bitstream is typically encrypted with the standard counter mode \cite{iso_mpeg}. 
The decoding process outputs different kinds of frames. Whereas I (intra-coded) frames are independent, the P (inter-predicted) frames are calculated using previous frames as a reference. B (bi-directional) frames use later frames as a reference, leading to out-of-order decoding. 
Therefore, multiple decoded frames are kept in off-chip memory buffers and if needed, are re-read by the inter-prediction stage.

\begin{figure}	
\begin{minipage}{\columnwidth}
	\centering
	\includegraphics[width=0.8\columnwidth]{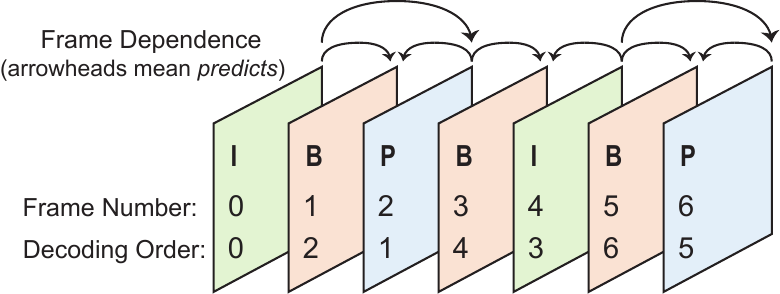}
	\vspace{0.2cm}
	\caption{H.264 decoding example involving predicted frames --- \small{I, P, and B represent independent, forward-predicted, and bidirectionally-predicted frames, respectively}.}
	\label{fig:h264-frames}
\end{minipage}
\end{figure}

\begin{figure}
	\begin{minipage}{\columnwidth}
		\centering
		\includegraphics[width=\columnwidth]{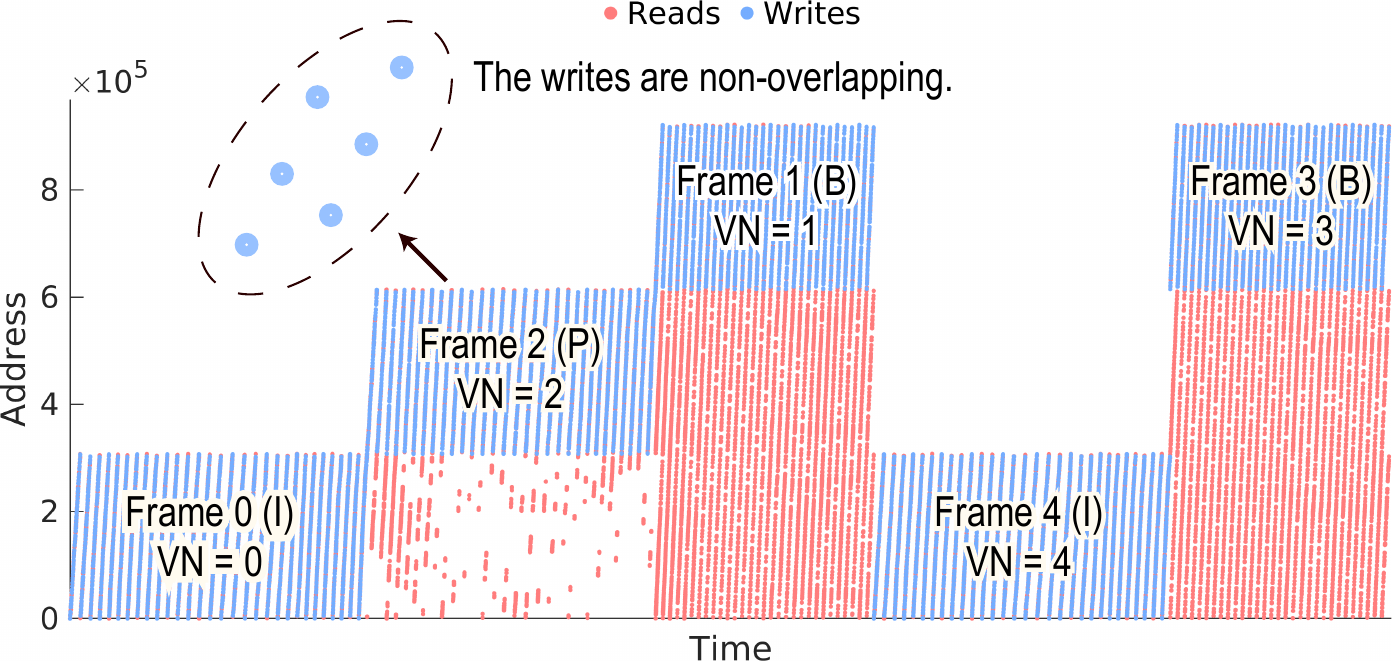}
		\caption{The memory access pattern of a H.264 decoder --- \small{the writes are non-overlapping}.}
		\vspace{0.2cm}
		\label{fig:h264-plot}
	\end{minipage}
	\vspace{-0.4cm}
\end{figure}

To study how MGX can be applied to a H.264 decoder, we analyzed an open-source implementation \cite{h264hls}. 
This decoder stores the decoded and reference frames in external memory, and supports the Main H.264 profile, 
which can have B frames. 
The decoder writes an output frame to an available buffer in external memory, but writes only once to an address 
in each frame. 
When a frame is used as a reference, it is read-only. Thus we can simply use the frame number ($F$)
concatenated with the input bitstream number (CTR$_\text{IN}$) as the VN when writing an output frame.
Both $F$ and CTR$_\text{IN}$ are part of the program state tracked by the scheduler.
CTR$_\text{IN}$ is incremented when a new bitstream is loaded for decoding.

The inter-prediction block can generate the VN for reading previously decoded frames based on the
current frame number ($F$). 
For the decoding of the IBPB sequence in Figure \ref{fig:h264-frames}, a P-frame is read only from the last I-frame, thus (CTR$_\text{IN}~||~F-2$) is used as the VN value.
Note that the frame number represents the display order of the frames, not the order of decoding. 
For decoding a B frame, frames from both directions are read; the VNs can be set to (CTR$_\text{IN}~||~F-1$) and (CTR$_\text{IN}~||~F+1$), respectively. 

We apply the MGX scheme to the H.264 decoder, performed an RTL simulation and checked functional correctness.
The memory access pattern is illustrated in Figure~\ref{fig:h264-plot} where there are three frame buffers in memory,
one for the currently decoded frame and two for reference frames.
The blue dots indicate writes and the pink dots indicate reads.
Because the frame number increments after writing a frame, our scheme ensures that a version number is different
for each write to a memory location.
While not clear from the figure due to a limited resolution, we verified that each location in the output buffer is 
written only once per frame.
The figure also shows that MGX can handle a dynamic and irregular read pattern.

\subsection{Static and Dynamic DNN Pruning}
Most previous pruning techniques prune a neural network statically~\cite{EIE, scp1, scp2, scp3},
which can simply be seen as a different network that can run on the secure DNN accelerator.
Dynamic pruning~\cite{scnn, cnvlutin, snapea, cgnet, cgnet_nips} exploits input-specific characteristics to skip redundant computations at run time, and
memory access patterns may vary for different inputs. 
However, the variations are still limited; dynamic pruning may skip some of the accesses 
that exist in the model, but does not introduce extra accesses.
It may appear that the MGX does not work with dynamic pruning.
However, skipping VNs does not affect the security as long as the VNs are not reused.
The memory protection will be functional as long as a write and the corresponding reads use the same VN.
\begin{figure}
\centering
\begin{minipage}{\linewidth}
  \centering
  \includegraphics[width=1.02\linewidth]{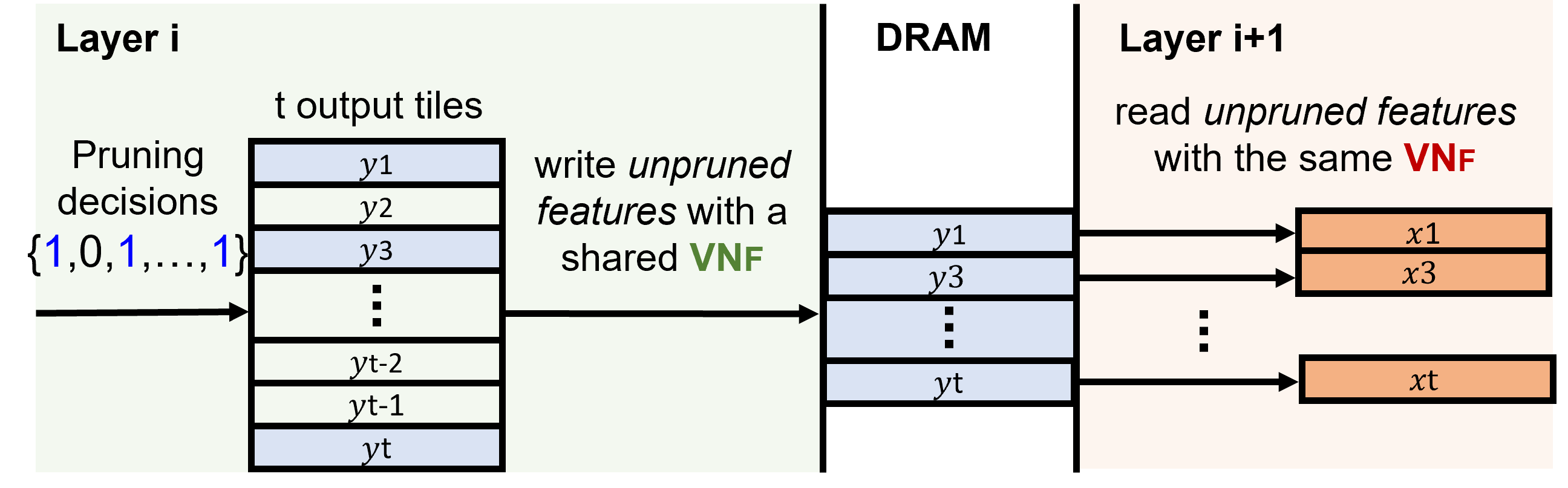}
  \captionof{figure}{A DNN layer with pruning writes only unpruned output tiles with a shared VN$_\text{F}$ value. The subsequent layers can therefore read the unpruned tiles with the same VN$_\text{F}$.}
  \label{fig:pruning}
\end{minipage}%
\end{figure}

To demonstrate the MGX under static and dynamic pruning, 
we implemented a variety of pruning techniques in PyTorch and emulated the MGX in software.
For pixel-level dynamic pruning, we implemented different compression techniques such as 
Compressed Sparse Row~\cite{csr}, Compressed Sparse Column~\cite{EIE,csc}, and 
Run-Length Compression~\cite{scnn}. We also tested a dynamic channel pruning scheme 
similar to~\cite{fbs}.
With dynamic pruning, the number of memory accesses to the features is input-dependent and determined at run time.
However, we found that the \versionnumbers~of the features with dynamic pruning can
still be obtained from the same VN generation scheme.
Figure~\ref{fig:pruning} illustrates the case where features are dynamically pruned. 
MGX uses the shared VN$_\text{F}$ to write output features, but only unpruned features (e.g., $x_1, x_3, ..., x_t$) are written to memory. 
Then, MGX can read the sparse input features (e.g., $x_1, x_3, ..., x_t$) using the same VN$_\text{F}$.
Again, only unpruned features are read from memory.
For pruned features, the VN is simply not used and skipped.
\color{black}
\section{Related Work}

\textbf{Privacy-Preserving Deep Learning --} Homomorphic encryption (HE) and secure multi-party computation (MPC) can provide stronger protection than TEEs by performing all computations in an encrypted format.
However, DNN tasks in the HE/MPC solutions \cite{CryptoNets_FHME, ONN_AHME, gazelle, wagh2019securenn, delphi, rathee2020cryptflow, cryptflow2020, wagh2020falcon, AriaNN} 
are still multiple orders of magnitude slower than the baseline with no protection.
A recent work~\cite{cheetah} proposes to reduce the latency of HE-based DNN inference to hundreds of milliseconds using specialized hardware.
Yet, the overhead in throughput is still quite significant even with an HE accelerator.
A secure accelerator with MGX provides a design point to offer hardware-based security with much higher performance.

There are many TEEs~\cite{sgx,trustzone,xom,aegis, secprocessor_micro03, secprocessor_isca05, ascend, secureME, hyperwall, CHERI, isox, secprocessor_hpca10, sanctum, mi6, keystone} proposed for CPUs.
Recent studies showed that DNNs can be protected using Intel SGX~\cite{privado, Vessels2020, Occlumency}, but with non-trivial overhead of memory protection in SGX. 
Today's processor-based TEEs are also limited by the performance of a general-purpose processor.
Recent studies~\cite{graviton, hix-asplos19, Telekine, htee-sp20} proposed to extend today's TEE by including a GPU. 
The GPU TEE designs enable high performance, but require 
both a CPU and a GPU to be protected inside a TEE.
Also, ASIC accelerators are often far more energy-efficient compared to GPUs and widely used for high-throughput tasks such as inference.
Outsourcing to untrusted GPUs/accelerators~\cite{slalom} is another promising approach. However, secure outsourcing introduces significant computation and storage overhead for the offline phase; for high-throughput applications, performance will still be limited by CPUs.

Recent work \cite{htee-sp20, NPUFort_secure_hardware, FPGA_TEE} proposes to build FPGA/ASIC TEEs as accelerators.
TNPU~\cite{TNPU} is most similar to our work, which also proposes a tree-free off-chip memory protection by exploiting the DNN-specific memory access patterns.
In this work, we demonstrate that MGX can further reduce the overhead of integrity verification using coarse-grained MACs and is generally applicable to other data-intensive accelerators beyond DNN accelerators.

\textbf{Memory Encryption and Integrity Verification --}
There is a large body of work on memory encryption and integrity verification for general-purpose CPUs,
including the counter-mode encryption~\cite{mem_enc}, optimizations to reduce the size of VNs~\cite{split_counter,morph},
counter-based integrity trees~\cite{iv_micro07,tec_tree, parallel_tree, vault},
meta-data caching~\cite{hash_tree, tree_cache}, and
predicting VNs or using unverified VNs~speculatively~\cite{counter_predict, ase, PoisonIvy}.
The general-purpose protection schemes all require version numbers in off-chip memory, which
will pose a challenge for applications with large data sets and random access patterns as in DLRM. 
MGX introduces a new approach to customize memory protection for a specific application
and remove off-chip VNs, 
which significantly reduces the overhead of the state-of-the-art.

\textbf{Side-channel Attacks and Protection --}
A variety of side-channel attacks have been shown against DNN accelerators. Memory and timing side-channels have been used to infer the network structure and weights of DNN models~\cite{recnn, recnn_journal, dnn_cache_side_channel}. 
Power and electromagnetic side-channel attacks have been used to retrieve the input image~\cite{powersidechan} or recover the network topology and weights~\cite{usenixem}.
The side channels are orthogonal to memory encryption and integrity verification that MGX aims to provide.
A secure accelerator needs to be extended with additional countermeasures to prevent the side channel attacks.
\section{Conclusion}
In this paper, we propose a novel off-chip memory protection scheme for hardware accelerators, named MGX. 
On average, MGX reduces the performance overhead of memory protection from 28\% and 33\% to 4\% and 5\% for DNN and graph processing accelerators, respectively.
We also show that MGX is generally applicable to other applications, such as genome assembly alignment and H.264 video decoding.
\section{Acknowledgment}
We thank the anonymous reviewers for their constructive feedback on the earlier version of the manuscript.
At Cornell, Weizhe Hua and Muhammad Umar are supported in part by NSF Award CCF-2007832, ECCS-1932501, and CCF-2118709. Weizhe Hua is also supported in part by the Facebook fellowship.

\bibliographystyle{ieeetr}
\bibliography{refs}

\begin{thebibliography}{10}

\bibitem{google2017tpu}
N.~P. Jouppi, C.~Young, N.~Patil, D.~Patterson, G.~Agrawal, R.~Bajwa, S.~Bates,
  S.~Bhatia, N.~Boden, A.~Borchers, R.~Boyle, P.-l. Cantin, C.~Chao, C.~Clark,
  J.~Coriell, M.~Daley, M.~Dau, J.~Dean, B.~Gelb, T.~V. Ghaemmaghami,
  R.~Gottipati, W.~Gulland, R.~Hagmann, C.~R. Ho, D.~Hogberg, J.~Hu, R.~Hundt,
  D.~Hurt, J.~Ibarz, A.~Jaffey, A.~Jaworski, A.~Kaplan, H.~Khaitan,
  D.~Killebrew, A.~Koch, N.~Kumar, S.~Lacy, J.~Laudon, J.~Law, D.~Le, C.~Leary,
  Z.~Liu, K.~Lucke, A.~Lundin, G.~MacKean, A.~Maggiore, M.~Mahony, K.~Miller,
  R.~Nagarajan, R.~Narayanaswami, R.~Ni, K.~Nix, T.~Norrie, M.~Omernick,
  N.~Penukonda, A.~Phelps, J.~Ross, M.~Ross, A.~Salek, E.~Samadiani, C.~Severn,
  G.~Sizikov, M.~Snelham, J.~Souter, D.~Steinberg, A.~Swing, M.~Tan,
  G.~Thorson, B.~Tian, H.~Toma, E.~Tuttle, V.~Vasudevan, R.~Walter, W.~Wang,
  E.~Wilcox, and D.~H. Yoon, ``In-datacenter performance analysis of a tensor
  processing unit,'' in {\em Proceedings of the 44th Annual International
  Symposium on Computer Architecture}, ISCA '17, (New York, NY, USA),
  p.~1–12, Association for Computing Machinery, 2017.

\bibitem{microsoft2018brainwave}
E.~Chung, J.~Fowers, K.~Ovtcharov, M.~Papamichael, A.~Caulfield, T.~Massengill,
  M.~Liu, {\em et~al.}, ``{Serving DNNs in Real Time at Datacenter Scale with
  Project Brainwave },'' {\em IEEE Micro}, vol.~38, no.~2, pp.~8--20, 2018.

\bibitem{sgx}
F.~McKeen, I.~Alexandrovich, I.~Anati, D.~Caspi, S.~Johnson, R.~Leslie-Hurd,
  and C.~Rozas, ``Intel® software guard extensions (intel® sgx) support for
  dynamic memory management inside an enclave,'' in {\em Proceedings of the
  Hardware and Architectural Support for Security and Privacy 2016}, HASP 2016,
  (New York, NY, USA), Association for Computing Machinery, 2016.

\bibitem{scale-sim1}
A.~Samajdar, Y.~Zhu, P.~Whatmough, M.~Mattina, and T.~Krishna, ``Scale-sim:
  Systolic cnn accelerator simulator,'' {\em arXiv preprint arXiv:1811.02883},
  2018.

\bibitem{graphlily}
Y.~Hu, Y.~Du, E.~Ustun, and Z.~Zhang, ``Graphlily: Accelerating graph linear
  algebra on hbm-equipped fpgas,'' {\em International Conference On Computer
  Aided Design}, 2021.

\bibitem{graviton}
S.~Volos, K.~Vaswani, and R.~Bruno, ``Graviton: Trusted execution environments
  on gpus,'' in {\em 13th {USENIX} Symposium on Operating Systems Design and
  Implementation ({OSDI} 18)}, (Carlsbad, CA), pp.~681--696, {USENIX}
  Association, Oct 2018.

\bibitem{encrypt_survey}
M.~Henson and S.~Taylor, ``Memory encryption: A survey of existing
  techniques,'' {\em ACM Comput. Surv.}, vol.~46, pp.~53:1--53:26, Mar 2014.

\bibitem{cpu_MEE}
S.~{Gueron}, ``Memory encryption for general-purpose processors,'' {\em IEEE
  Security Privacy}, vol.~14, pp.~54--62, Nov 2016.

\bibitem{aegis}
G.~E. Suh, D.~Clarke, B.~Gassend, M.~van Dijk, and S.~Devadas, ``{AEGIS:
  Architecture for Tamper-evident and Tamper-resistant Processing},'' in {\em
  Proceedings of the 17th Annual International Conference on Supercomputing},
  ICS '03, (New York, NY, USA), pp.~160--171, ACM, 2003.

\bibitem{hash_tree}
B.~Gassend, G.~Suh, D.~Clarke, M.~van Dijk, and S.~Devadas, ``Caches and hash
  trees for efficient memory integrity verification,'' in {\em The Ninth
  International Symposium on High-Performance Computer Architecture, 2003.
  HPCA-9 2003. Proceedings.}, pp.~295--306, 2003.

\bibitem{iv_micro07}
B.~Rogers, S.~Chhabra, M.~Prvulovic, and Y.~Solihin, ``{Using Address
  Independent Seed Encryption and Bonsai Merkle Trees to Make Secure Processors
  OS- and Performance-Friendly},'' in {\em Proceedings of the 40th Annual
  IEEE/ACM International Symposium on Microarchitecture}, MICRO 40,
  (Washington, DC, USA), pp.~183--196, IEEE Computer Society, 2007.

\bibitem{tensorflow}
M.~Abadi, P.~Barham, J.~Chen, Z.~Chen, A.~Davis, J.~Dean, M.~Devin,
  S.~Ghemawat, G.~Irving, M.~Isard, M.~Kudlur, J.~Levenberg, R.~Monga,
  S.~Moore, D.~G. Murray, B.~Steiner, P.~Tucker, V.~Vasudevan, P.~Warden,
  M.~Wicke, Y.~Yu, and X.~Zheng, ``Tensorflow: A system for large-scale machine
  learning,'' in {\em Proceedings of the 12th USENIX Conference on Operating
  Systems Design and Implementation}, OSDI'16, (Berkeley, CA, USA),
  pp.~265--283, USENIX Association, 2016.

\bibitem{chaidnn}
Xilinx, ``{CHaiDNN-v2: HLS based Deep Neural Network Accelerator Library for
  Xilinx Ultrascale+ MPSoCs}.'' \url{https://github.com/Xilinx/CHaiDNN}, Jun
  2018.

\bibitem{aes_nist}
M.~J. Dworkin, ``Sp 800-38c. recommendation for block cipher modes of
  operation: The ccm mode for authentication and confidentiality,'' tech. rep.,
  Gaithersburg, MD, USA, 2004.

\bibitem{aes_comment}
H.~Lipmaa, D.~Wagner, and P.~Rogaway, ``Comments to nist concerning aes modes
  of operation: Ctr-mode encryption,'' 2000.

\bibitem{hash_tree_proof}
R.~C. Merkle, ``Protocols for public key cryptosystems,'' in {\em 1980 IEEE
  Symposium on Security and Privacy}, pp.~122--122, 1980.

\bibitem{glow}
N.~Rotem, J.~Fix, S.~Abdulrasool, S.~Deng, R.~Dzhabarov, J.~Hegeman,
  R.~Levenstein, B.~Maher, N.~Satish, J.~Olesen, J.~Park, A.~Rakhov, and
  M.~Smelyanskiy, ``Glow: Graph lowering compiler techniques for neural
  networks,'' {\em CoRR}, vol.~abs/1805.00907, 2018.

\bibitem{tvm}
T.~Chen, T.~Moreau, Z.~Jiang, L.~Zheng, E.~Yan, H.~Shen, M.~Cowan, L.~Wang,
  Y.~Hu, L.~Ceze, C.~Guestrin, and A.~Krishnamurthy, ``{TVM}: An automated
  end-to-end optimizing compiler for deep learning,'' in {\em 13th {USENIX}
  Symposium on Operating Systems Design and Implementation ({OSDI} 18)},
  (Carlsbad, CA), pp.~578--594, {USENIX} Association, Oct 2018.

\bibitem{vta}
T.~Moreau, T.~Chen, L.~Vega, J.~Roesch, E.~Yan, L.~Zheng, J.~Fromm, Z.~Jiang,
  L.~Ceze, C.~Guestrin, and A.~Krishnamurthy, ``A hardware–software blueprint
  for flexible deep learning specialization,'' {\em IEEE Micro}, vol.~39,
  no.~5, pp.~8--16, 2019.

\bibitem{he2015resnet}
K.~He, X.~Zhang, S.~Ren, and J.~Sun, ``Deep residual learning for image
  recognition,'' in {\em 2016 IEEE Conference on Computer Vision and Pattern
  Recognition (CVPR)}, pp.~770--778, 2016.

\bibitem{howard2017mobilenets}
A.~G. Howard, M.~Zhu, B.~Chen, D.~Kalenichenko, W.~Wang, T.~Weyand,
  M.~Andreetto, and H.~Adam, ``{Mobilenets: Efficient Convolutional Neural
  Networks for Nobile Vision Applications},'' {\em arXiv e-print},
  vol.~arXiv:1704.04861, 2017.

\bibitem{graph_blas}
J.~Kepner, P.~Aaltonen, D.~Bader, A.~Buluc, F.~Franchetti, J.~Gilbert,
  D.~Hutchison, M.~Kumar, A.~Lumsdaine, H.~Meyerhenke, S.~McMillan, C.~Yang,
  J.~D. Owens, M.~Zalewski, T.~Mattson, and J.~Moreira, ``Mathematical
  foundations of the graphblas,'' {\em 2016 IEEE High Performance Extreme
  Computing Conference, HPEC 2016}, 12 2016.

\bibitem{pagerank_fpga}
S.~Zhou, C.~Chelmis, and V.~K. Prasanna, ``Optimizing memory performance for
  fpga implementation of pagerank,'' in {\em 2015 International Conference on
  ReConFigurable Computing and FPGAs (ReConFig)}, pp.~1--6, 2015.

\bibitem{bfs_fpga}
J.~Zhang, S.~Khoram, and J.~Li, ``Boosting the performance of fpga-based graph
  processor using hybrid memory cube: A case for breadth first search,'' in
  {\em Proceedings of the 2017 ACM/SIGDA International Symposium on
  Field-Programmable Gate Arrays}, FPGA '17, (New York, NY, USA), p.~207–216,
  Association for Computing Machinery, 2017.

\bibitem{graph-dataset1}
T.~A. Davis and Y.~Hu, ``The university of florida sparse matrix collection,''
  {\em ACM Trans. Math. Softw.}, vol.~38, Dec 2011.

\bibitem{graph-dataset2}
H.~Kwak, C.~Lee, H.~Park, and S.~Moon, ``What is twitter, a social network or a
  news media?,'' in {\em Proceedings of the 19th International Conference on
  World Wide Web}, WWW ’10, (New York, NY, USA), p.~591–600, Association
  for Computing Machinery, 2010.

\bibitem{hu2020ogb}
W.~Hu, M.~Fey, M.~Zitnik, Y.~Dong, H.~Ren, B.~Liu, M.~Catasta, and J.~Leskovec,
  ``Open graph benchmark: Datasets for machine learning on graphs,'' {\em arXiv
  preprint arXiv:2005.00687}, 2020.

\bibitem{ramulator}
Y.~Kim, W.~Yang, and O.~Mutlu, ``Ramulator: A fast and extensible dram
  simulator,'' {\em IEEE CAL}, vol.~15, no.~1, pp.~45--49, 2016.

\bibitem{samsung-npu}
Y.~Duk~Kim, W.~Jeong, L.~Jung, D.~Shin, J.~G. Song, J.~Song, H.~Kwon, J.~Lee,
  J.~Jung, M.~Kang, J.~Jeong, Y.~Kwon, and N.~H. Seong, ``2.4 a 7nm
  high-performance and energy-efficient mobile application processor with
  tri-cluster cpus and a sparsity-aware npu,'' in {\em 2020 IEEE International
  Solid- State Circuits Conference - (ISSCC)}, pp.~48--50, 2020.

\bibitem{snapnets}
J.~Leskovec and A.~Krevl, ``{SNAP Datasets}: {Stanford} large network dataset
  collection.'' \url{http://snap.stanford.edu/data}, Jun 2014.

\bibitem{aes_gcm}
S.~Lemsitzer, J.~Wolkerstorfer, N.~Felber, and M.~Braendli, ``Multi-gigabit
  gcm-aes architecture optimized for fpgas,'' in {\em Cryptographic Hardware
  and Embedded Systems - CHES 2007} (P.~Paillier and I.~Verbauwhede, eds.),
  (Berlin, Heidelberg), pp.~227--238, Springer Berlin Heidelberg, 2007.

\bibitem{darwin}
Y.~Turakhia, G.~Bejerano, and W.~J. Dally, ``Darwin: A genomics co-processor
  provides up to 15,000x acceleration on long read assembly,'' in {\em
  Proceedings of the Twenty-Third International Conference on Architectural
  Support for Programming Languages and Operating Systems}, ASPLOS ’18, (New
  York, NY, USA), p.~199–213, Association for Computing Machinery, 2018.

\bibitem{grch38}
G.~R. Consortium, ``{Genome Reference Consortium Human Build 38}.''
  \url{https://www.ncbi.nlm.nih.gov/assembly/GCF_000001405.26/}, Dec 2013.

\bibitem{iso_h264avc}
``{Information technology — Coding of audio-visual objects — Part 10:
  Advanced Video Coding},'' Standard ISO/IEC 14496-10:2014, International
  Organization for Standardization, Geneva, CH, 2014.

\bibitem{iso_mpeg}
``{Information technology — MPEG systems technologies — Part 7: Common
  encryption in ISO base media file format files},'' Standard ISO/IEC
  23001-7:2016, International Organization for Standardization, Geneva, CH,
  2016.

\bibitem{h264hls}
X.~Liu, Y.~Chen, T.~Nguyen, S.~Gurumani, K.~Rupnow, and D.~Chen, ``High level
  synthesis of complex applications: An h.264 video decoder,'' in {\em
  Proceedings of the 2016 ACM/SIGDA International Symposium on
  Field-Programmable Gate Arrays}, FPGA ’16, (New York, NY, USA),
  p.~224–233, Association for Computing Machinery, 2016.

\bibitem{EIE}
S.~Han, X.~Liu, H.~Mao, J.~Pu, A.~Pedram, M.~A. Horowitz, and W.~J. Dally,
  ``Eie: Efficient inference engine on compressed deep neural network,'' in
  {\em Proceedings of the 43rd International Symposium on Computer
  Architecture}, ISCA '16, p.~243–254, IEEE Press, 2016.

\bibitem{scp1}
H.~Li, A.~Kadav, I.~Durdanovic, H.~Samet, and H.~P. Graf, ``{Pruning Filters
  for Efficient ConvNets},'' in {\em International Conference on Learning
  Representations}, May 2017.

\bibitem{scp2}
Z.~Liu, J.~Li, Z.~Shen, G.~Huang, S.~Yan, and C.~Zhang, ``Learning efficient
  convolutional networks through network slimming,'' in {\em ICCV}, 2017.

\bibitem{scp3}
Y.~He, X.~Zhang, and J.~Sun, ``Channel pruning for accelerating very deep
  neural networks,'' in {\em The IEEE International Conference on Computer
  Vision (ICCV)}, Oct 2017.

\bibitem{scnn}
A.~Parashar, M.~Rhu, A.~Mukkara, A.~Puglielli, R.~Venkatesan, B.~Khailany,
  J.~Emer, S.~W. Keckler, and W.~J. Dally, ``Scnn: An accelerator for
  compressed-sparse convolutional neural networks,'' in {\em 2017 ACM/IEEE 44th
  Annual International Symposium on Computer Architecture (ISCA)}, pp.~27--40,
  2017.

\bibitem{cnvlutin}
J.~Albericio, P.~Judd, T.~Hetherington, T.~Aamodt, N.~E. Jerger, and
  A.~Moshovos, ``Cnvlutin: Ineffectual-neuron-free deep neural network
  computing,'' in {\em 2016 ACM/IEEE 43rd Annual International Symposium on
  Computer Architecture (ISCA)}, pp.~1--13, 2016.

\bibitem{snapea}
V.~Akhlaghi, A.~Yazdanbakhsh, K.~Samadi, R.~K. Gupta, and H.~Esmaeilzadeh,
  ``Snapea: Predictive early activation for reducing computation in deep
  convolutional neural networks,'' in {\em 2018 ACM/IEEE 45th Annual
  International Symposium on Computer Architecture (ISCA)}, pp.~662--673, 2018.

\bibitem{cgnet}
W.~Hua, Y.~Zhou, C.~De~Sa, Z.~Zhang, and G.~E. Suh, ``{Boosting the Performance
  of CNN Accelerators with Dynamic Fine-Grained Channel Gating},'' in {\em
  Proceedings of the 52Nd Annual IEEE/ACM International Symposium on
  Microarchitecture}, MICRO '52, (New York, NY, USA), pp.~139--150, ACM, 2019.

\bibitem{cgnet_nips}
W.~Hua, Y.~Zhou, C.~M. De~Sa, Z.~Zhang, and G.~E. Suh, ``{Channel Gating Neural
  Networks},'' in {\em Advances in Neural Information Processing Systems 32}
  (H.~Wallach, H.~Larochelle, A.~Beygelzimer, F.~d\textquotesingle
  Alch\'{e}-Buc, E.~Fox, and R.~Garnett, eds.), pp.~1884--1894, Curran
  Associates, Inc., 2019.

\bibitem{csr}
J.~{Albericio}, P.~{Judd}, T.~{Hetherington}, T.~{Aamodt}, N.~E. {Jerger}, and
  A.~{Moshovos}, ``Cnvlutin: Ineffectual-neuron-free deep neural network
  computing,'' in {\em 2016 ACM/IEEE 43rd Annual International Symposium on
  Computer Architecture (ISCA)}, pp.~1--13, 2016.

\bibitem{csc}
V.~Sze, Y.~Chen, T.~Yang, and J.~S. Emer, ``Efficient processing of deep neural
  networks: {A} tutorial and survey,'' {\em CoRR}, vol.~abs/1703.09039, 2017.

\bibitem{fbs}
X.~Gao, Y.~Zhao, Łukasz Dudziak, R.~Mullins, and C.~zhong Xu, ``Dynamic
  channel pruning: Feature boosting and suppression,'' in {\em International
  Conference on Learning Representations}, 2019.

\bibitem{CryptoNets_FHME}
N.~Dowlin, R.~Gilad-Bachrach, K.~Laine, K.~Lauter, M.~Naehrig, and J.~Wernsing,
  ``Cryptonets: Applying neural networks to encrypted data with high throughput
  and accuracy,'' in {\em Proceedings of the 33rd International Conference on
  International Conference on Machine Learning - Volume 48}, ICML'16,
  p.~201–210, JMLR.org, 2016.

\bibitem{ONN_AHME}
J.~Liu, M.~Juuti, Y.~Lu, and N.~Asokan, ``Oblivious neural network predictions
  via minionn transformations,'' in {\em Proceedings of the 2017 ACM SIGSAC
  Conference on Computer and Communications Security}, CCS '17, (New York, NY,
  USA), p.~619–631, Association for Computing Machinery, 2017.

\bibitem{gazelle}
C.~Juvekar, V.~Vaikuntanathan, and A.~Chandrakasan, ``Gazelle: A low latency
  framework for secure neural network inference,'' in {\em Proceedings of the
  27th USENIX Conference on Security Symposium}, SEC’18, (USA),
  p.~1651–1668, USENIX Association, 2018.

\bibitem{wagh2019securenn}
S.~Wagh, D.~Gupta, and N.~Chandran, ``Securenn: Efficient and private neural
  network training,'' in {\em Privacy Enhancing Technologies Symposium}, (PETS
  2019), February 2019.

\bibitem{delphi}
P.~Mishra, R.~Lehmkuhl, A.~Srinivasan, W.~Zheng, and R.~A. Popa, ``Delphi: A
  cryptographic inference service for neural networks,'' in {\em 29th {USENIX}
  Security Symposium ({USENIX} Security 20)}, (Boston, MA), {USENIX}
  Association, Aug 2020.

\bibitem{rathee2020cryptflow}
D.~Rathee, M.~Rathee, N.~Kumar, N.~Chandran, D.~Gupta, A.~Rastogi, and
  R.~Sharma, ``Cryptflow2: Practical 2-party secure inference,'' in {\em 27th
  Annual Conference on Computer and Communications Security (ACM CCS 2020)},
  ACM, August 2020.

\bibitem{cryptflow2020}
N.~Kumar, M.~Rathee, N.~Chandran, D.~Gupta, A.~Rastogi, and R.~Sharma,
  ``Cryptflow: Secure tensorflow inference,'' in {\em 2020 IEEE Symposium on
  Security and Privacy (SP)}, pp.~336--353, 2020.

\bibitem{wagh2020falcon}
S.~Wagh, S.~Tople, F.~Benhamouda, E.~Kushilevitz, P.~Mittal, and T.~Rabin,
  ``{FALCON: Honest-Majority Maliciously Secure Framework for Private Deep
  Learning},'' 2021.

\bibitem{AriaNN}
T.~Ryffel, P.~Tholoniat, D.~Pointcheval, and F.~Bach, ``Ariann: Low-interaction
  privacy-preserving deep learning via function secret sharing,'' {\em
  Proceedings on Privacy Enhancing Technologies}, vol.~2022, no.~1,
  pp.~291--316, 2022.

\bibitem{cheetah}
B.~Reagen, W.-S. Choi, Y.~Ko, V.~T. Lee, H.-H.~S. Lee, G.-Y. Wei, and
  D.~Brooks, ``Cheetah: Optimizing and accelerating homomorphic encryption for
  private inference,'' in {\em 2021 IEEE International Symposium on
  High-Performance Computer Architecture (HPCA)}, pp.~26--39, 2021.

\bibitem{trustzone}
T.~Alves and D.~Felton, ``Trustzone: Integrated hardware and software
  security,'' 01 2004.

\bibitem{xom}
D.~L.~C. Thekkath, M.~Mitchell, P.~Lincoln, D.~Boneh, J.~Mitchell, and
  M.~Horowitz, ``Architectural support for copy and tamper resistant
  software,'' in {\em Proceedings of the Ninth International Conference on
  Architectural Support for Programming Languages and Operating Systems},
  ASPLOS IX, (New York, NY, USA), pp.~168--177, ACM, 2000.

\bibitem{secprocessor_micro03}
J.~Yang, Y.~Zhang, and L.~Gao, ``Fast secure processor for inhibiting software
  piracy and tampering,'' in {\em Proceedings of the 36th Annual IEEE/ACM
  International Symposium on Microarchitecture}, MICRO 36, (USA), p.~351, IEEE
  Computer Society, 2003.

\bibitem{secprocessor_isca05}
R.~B. Lee, P.~C.~S. Kwan, J.~P. McGregor, J.~Dwoskin, and Z.~Wang,
  ``Architecture for protecting critical secrets in microprocessors,'' in {\em
  Proceedings of the 32nd Annual International Symposium on Computer
  Architecture}, ISCA '05, (USA), p.~2–13, IEEE Computer Society, 2005.

\bibitem{ascend}
C.~W. Fletcher, M.~v. Dijk, and S.~Devadas, ``A secure processor architecture
  for encrypted computation on untrusted programs,'' in {\em Proceedings of the
  Seventh ACM Workshop on Scalable Trusted Computing}, STC '12, (New York, NY,
  USA), pp.~3--8, ACM, 2012.

\bibitem{secureME}
S.~Chhabra, B.~Rogers, Y.~Solihin, and M.~Prvulovic, ``Secureme: A
  hardware-software approach to full system security,'' in {\em Proceedings of
  the International Conference on Supercomputing}, ICS '11, (New York, NY,
  USA), p.~108–119, Association for Computing Machinery, 2011.

\bibitem{hyperwall}
J.~Szefer and R.~B. Lee, ``Architectural support for hypervisor-secure
  virtualization,'' in {\em Proceedings of the Seventeenth International
  Conference on Architectural Support for Programming Languages and Operating
  Systems}, ASPLOS XVII, (New York, NY, USA), p.~437–450, Association for
  Computing Machinery, 2012.

\bibitem{CHERI}
R.~N. Watson, J.~Woodruff, P.~G. Neumann, S.~W. Moore, J.~Anderson,
  D.~Chisnall, N.~Dave, B.~Davis, K.~Gudka, B.~Laurie, S.~J. Murdoch,
  R.~Norton, M.~Roe, S.~Son, and M.~Vadera, ``Cheri: A hybrid capability-system
  architecture for scalable software compartmentalization,'' in {\em 2015 IEEE
  Symposium on Security and Privacy}, pp.~20--37, 2015.

\bibitem{isox}
D.~{Evtyushkin}, J.~{Elwell}, M.~{Ozsoy}, D.~{Ponomarev}, N.~A. {Ghazaleh}, and
  R.~{Riley}, ``Iso-x: A flexible architecture for hardware-managed isolated
  execution,'' in {\em 2014 47th Annual IEEE/ACM International Symposium on
  Microarchitecture}, pp.~190--202, 2014.

\bibitem{secprocessor_hpca10}
D.~Champagne and R.~B. Lee, ``Scalable architectural support for trusted
  software,'' in {\em HPCA - 16 2010 The Sixteenth International Symposium on
  High-Performance Computer Architecture}, pp.~1--12, 2010.

\bibitem{sanctum}
V.~Costan, I.~Lebedev, and S.~Devadas, ``Sanctum: Minimal hardware extensions
  for strong software isolation,'' in {\em 25th {USENIX} Security Symposium
  ({USENIX} Security 16)}, (Austin, TX), pp.~857--874, {USENIX} Association,
  Aug 2016.

\bibitem{mi6}
T.~Bourgeat, I.~Lebedev, A.~Wright, S.~Zhang, Arvind, and S.~Devadas, ``Mi6:
  Secure enclaves in a speculative out-of-order processor,'' in {\em
  Proceedings of the 52nd Annual IEEE/ACM International Symposium on
  Microarchitecture}, MICRO '52, (New York, NY, USA), p.~42–56, Association
  for Computing Machinery, 2019.

\bibitem{keystone}
D.~Lee, D.~Kohlbrenner, S.~Shinde, K.~Asanovic, and D.~Song, ``Keystone: An
  open framework for architecting trusted execution environments,'' in {\em
  Proceedings of the Fifteenth European Conference on Computer Systems},
  EuroSys’20, 2020.

\bibitem{privado}
S.~Tople, K.~Grover, S.~Shinde, R.~Bhagwan, and R.~Ramjee, ``Privado: Practical
  and secure {DNN} inference,'' {\em CoRR}, vol.~abs/1810.00602, 2018.

\bibitem{Vessels2020}
K.~Kim, C.~H. Kim, J.~J. Rhee, X.~Yu, H.~Chen, D.~J. Tian, and B.~Lee,
  ``Vessels: Efficient and scalable deep learning prediction on trusted
  processors,'' in {\em Proceedings of the 11th ACM Symposium on Cloud
  Computing}, SoCC '20, (New York, NY, USA), p.~462–476, Association for
  Computing Machinery, 2020.

\bibitem{Occlumency}
T.~Lee, Z.~Lin, S.~Pushp, C.~Li, Y.~Liu, Y.~Lee, F.~Xu, C.~Xu, L.~Zhang, and
  J.~Song, ``Occlumency: Privacy-preserving remote deep-learning inference
  using sgx,'' in {\em The 25th Annual International Conference on Mobile
  Computing and Networking}, MobiCom '19, (New York, NY, USA), Association for
  Computing Machinery, 2019.

\bibitem{hix-asplos19}
I.~Jang, A.~Tang, T.~Kim, S.~Sethumadhavan, and J.~Huh, ``Heterogeneous
  isolated execution for commodity gpus,'' in {\em Proceedings of the
  Twenty-Fourth International Conference on Architectural Support for
  Programming Languages and Operating Systems}, ASPLOS '19, (New York, NY,
  USA), p.~455–468, Association for Computing Machinery, 2019.

\bibitem{Telekine}
T.~Hunt, Z.~Jia, V.~Miller, A.~Szekely, Y.~Hu, C.~J. Rossbach, and E.~Witchel,
  ``Telekine: Secure computing with cloud {GPUs},'' in {\em 17th USENIX
  Symposium on Networked Systems Design and Implementation (NSDI 20)}, (Santa
  Clara, CA), pp.~817--833, USENIX Association, Feb 2020.

\bibitem{htee-sp20}
J.~Zhu, R.~Hou, X.~Wang, W.~Wang, J.~Cao, B.~Zhao, Z.~Wang, Y.~Zhang, J.~Ying,
  L.~Zhang, and D.~Meng, ``Enabling rack-scale confidential computing using
  heterogeneous trusted execution environment,'' in {\em 2020 IEEE Symposium on
  Security and Privacy (SP)}, pp.~1450--1465, 2020.

\bibitem{slalom}
F.~Tramer and D.~Boneh, ``Slalom: Fast, verifiable and private execution of
  neural networks in trusted hardware,'' in {\em International Conference on
  Learning Representations}, 2019.

\bibitem{NPUFort_secure_hardware}
X.~Wang, R.~Hou, Y.~Zhu, J.~Zhang, and D.~Meng, ``Npufort: A secure
  architecture of dnn accelerator against model inversion attack,'' in {\em
  Proceedings of the 16th ACM International Conference on Computing Frontiers},
  CF '19, (New York, NY, USA), pp.~190--196, ACM, 2019.

\bibitem{FPGA_TEE}
M.~Zhao, M.~Gao, and C.~Kozyrakis, {\em ShEF: Shielded Enclaves for Cloud
  FPGAs}, p.~1070–1085.
\newblock New York, NY, USA: Association for Computing Machinery, 2022.

\bibitem{TNPU}
S.~Lee, J.~Kim, S.~Na, J.~Park, and J.~Huh, ``Tnpu: Supporting trusted
  execution with tree-less integrity protection for neural processing unit,''
  in {\em 2022 IEEE International Symposium on High-Performance Computer
  Architecture (HPCA)}, pp.~229--243, 2022.

\bibitem{mem_enc}
G.~E. Suh, D.~Clarke, B.~Gassend, M.~v. Dijk, and S.~Devadas, ``Efficient
  memory integrity verification and encryption for secure processors,'' in {\em
  Proceedings of the 36th Annual IEEE/ACM International Symposium on
  Microarchitecture}, MICRO 36, (Washington, DC, USA), pp.~339--, IEEE Computer
  Society, 2003.

\bibitem{split_counter}
C.~Yan, D.~Englender, M.~Prvulovic, B.~Rogers, and Y.~Solihin, ``Improving
  cost, performance, and security of memory encryption and authentication,''
  {\em SIGARCH Comput. Archit. News}, vol.~34, pp.~179--190, May 2006.

\bibitem{morph}
G.~Saileshwar, P.~J. Nair, P.~Ramrakhyani, W.~Elsasser, J.~A. Joao, and M.~K.
  Qureshi, ``Morphable counters: Enabling compact integrity trees for
  low-overhead secure memories,'' in {\em 2018 51st Annual IEEE/ACM
  International Symposium on Microarchitecture (MICRO)}, pp.~416--427, Oct
  2018.

\bibitem{tec_tree}
R.~Elbaz, D.~Champagne, R.~B. Lee, L.~Torres, G.~Sassatelli, and P.~Guillemin,
  ``Tec-tree: A low-cost, parallelizable tree for efficient defense against
  memory replay attacks,'' in {\em Cryptographic Hardware and Embedded Systems
  - CHES 2007} (P.~Paillier and I.~Verbauwhede, eds.), (Berlin, Heidelberg),
  pp.~289--302, Springer Berlin Heidelberg, 2007.

\bibitem{parallel_tree}
W.~E. Hall and C.~S. Jutla, ``Parallelizable authentication trees,'' in {\em
  Proceedings of the 12th International Conference on Selected Areas in
  Cryptography}, SAC'05, (Berlin, Heidelberg), pp.~95--109, Springer-Verlag,
  2006.

\bibitem{vault}
M.~Taassori, A.~Shafiee, and R.~Balasubramonian, ``Vault: Reducing paging
  overheads in sgx with efficient integrity verification structures,'' in {\em
  Proceedings of the Twenty-Third International Conference on Architectural
  Support for Programming Languages and Operating Systems}, ASPLOS '18, (New
  York, NY, USA), pp.~665--678, ACM, 2018.

\bibitem{tree_cache}
J.~Lee, T.~Kim, and J.~Huh, ``Reducing the memory bandwidth overheads of
  hardware security support for multi-core processors,'' {\em IEEE Transactions
  on Computers}, vol.~65, pp.~3384--3397, Nov 2016.

\bibitem{counter_predict}
{Weidong Shi}, H.~S. {Lee}, M.~{Ghosh}, {Chenghuai Lu}, and A.~{Boldyreva},
  ``High efficiency counter mode security architecture via prediction and
  precomputation,'' in {\em 32nd International Symposium on Computer
  Architecture (ISCA'05)}, pp.~14--24, June 2005.

\bibitem{ase}
W.~Shi and H.-H.~S. Lee, ``ase,'' in {\em Proceedings of the 39th Annual
  IEEE/ACM International Symposium on Microarchitecture}, MICRO 39,
  (Washington, DC, USA), pp.~103--112, IEEE Computer Society, 2006.

\bibitem{PoisonIvy}
T.~S. Lehman, A.~D. Hilton, and B.~C. Lee, ``Poisonivy: Safe speculation for
  secure memory,'' in {\em 2016 49th Annual IEEE/ACM International Symposium on
  Microarchitecture (MICRO)}, pp.~1--13, Oct 2016.

\bibitem{recnn}
W.~Hua, Z.~Zhang, and G.~E. Suh, ``Reverse engineering convolutional neural
  networks through side-channel information leaks,'' in {\em Proceedings of the
  55th Annual Design Automation Conference}, DAC '18, (New York, NY, USA),
  pp.~4:1--4:6, ACM, 2018.

\bibitem{recnn_journal}
W.~Hua, Z.~Zhang, and G.~Edward~Suh, ``Reverse engineering cnn models using
  side-channel attacks,'' {\em IEEE Design Test}, 2022.

\bibitem{dnn_cache_side_channel}
M.~Yan, C.~W. Fletcher, and J.~Torrellas, ``Cache telepathy: Leveraging shared
  resource attacks to learn {DNN} architectures,'' in {\em 29th USENIX Security
  Symposium (USENIX Security 20)}, pp.~2003--2020, USENIX Association, Aug
  2020.

\bibitem{powersidechan}
L.~Wei, B.~Luo, Y.~Li, Y.~Liu, and Q.~Xu, ``I know what you see: Power
  side-channel attack on convolutional neural network accelerators,'' in {\em
  Proceedings of the 34th Annual Computer Security Applications Conference},
  ACSAC '18, (New York, NY, USA), pp.~393--406, ACM, 2018.

\bibitem{usenixem}
L.~Batina, S.~Bhasin, D.~Jap, and S.~Picek, ``{CSI} {NN}: Reverse engineering
  of neural network architectures through electromagnetic side channel,'' in
  {\em 28th {USENIX} Security Symposium ({USENIX} Security 19)}, (Santa Clara,
  CA), pp.~515--532, {USENIX} Association, Aug 2019.

\end{thebibliography}

\end{document}